\documentclass[apj,twocolumn,twocolappendix,numberedappendix]{openjournal}
\usepackage{amsmath}
\usepackage{booktabs}
\usepackage{multirow}
\usepackage{color}
\usepackage{soul}
\usepackage{threeparttable}
\usepackage[utf8]{inputenc}
\usepackage{float}
\usepackage{graphicx}
\usepackage{CJK}
\usepackage{xspace}
\usepackage{afterpage}
\usepackage{placeins}
\usepackage[breaklinks,colorlinks,citecolor=blue,urlcolor=blue,linkcolor=blue,filecolor=blue]{hyperref}

\shorttitle{JWST's GLIMPSE}
\shortauthors{}

\usepackage[export]{adjustbox}
\newcommand{\ha}{H$\alpha$}

\newcommand{\orcid}[1]{\href{https://orcid.org/#1}{\textcolor[HTML]{A6CE39}{\aiOrcid}}}

\def\ltsima{$\buildrel<\over\sim$}
\def\lsim{\lower.5ex\hbox{\ltsima}~}
\def\gtsima{$\buildrel>\over\sim$}
\def\gsim{\lower.5ex\hbox{\gtsima}~}

\newcommand{\muv}{$M_{\mathrm {UV}}$}

\def\msol{M$_{\odot}$}
\def\mstar{M$_{\star}$}

\def\ha{H$\alpha$}

\def\aa{\AA~}

\newcommand{\fesc}{$f_{\mathrm{esc}}$}

\def\cm2{cm$^{-2}$}

\def\ewo3{$EW_{\mathrm{[O\textsc{iii}]}}$}

\def\nh{\ifmmode N_{\mathrm{HI}}\else $N_{\mathrm{HI}}$\fi}

\def\xiion{$\xi_{\mathrm{ion}}$}

\newcommand{\glimpse}{{\sc Glimpse}}
\newcommand{\jwst}{{\em JWST}}
\newcommand{\hst}{{\em HST}}
\newcommand{\ra}[3]{#1$^{\mathrm{h}}$#2$^{\mathrm{m}}$#3$^{\mathrm{s}}$}
\newcommand{\decl}[3]
{#1$^{\circ}$#2$'$#3$''$}

\begin{document}

\title[]{JWST's GLIMPSE: an overview of the deepest probe of early galaxy formation and cosmic reionization\vspace{-1.5cm}}

\author{Hakim Atek (\includegraphics[height=1.7ex]{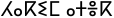})$^{1*}$, John Chisholm$^{2,3}$, Vasily Kokorev$^{2,3}$, Ryan Endsley$^{2,3}$, Richard Pan$^{4}$, Lukas J. Furtak $^{2,3}$, Iryna Chemerynska$^{1}$, Johan Richard$^{5}$, Ad\'ela\"ide Claeyssens$^{5}$, Pascal A. Oesch$^{6,7}$, Seiji Fujimoto$^{8,9}$, Rohan P. Naidu$^{10}$, Damien Korber$^{6}$, Daniel Schaerer$^{6,11}$, Jeremy Blaizot$^{5}$, Joki Rosdahl$^{5}$, Angela Adamo$^{12}$, Yoshihisa Asada$^{9}$, Arghyadeep Basu$^{5}$, Benjamin Beauchesne$^{13,14}$, Danielle Berg$^{2,3}$, Rachel Bezanson$^{15}$, Rychard Bouwens$^{16}$, Gabriel Brammer$^{7}$, Miroslava Dessauges-Zavadsky$^{6}$, Ama\"el Ellien$^{17}$, Meriam Ezziati$^{1}$, Qinyue Fei$^{8}$, Ilias Goovaerts$^{18}$, Sylvain Heurtier$^{1}$, Tiger Yu-Yang Hsiao$^{2,3}$, Michelle Jecmen$^{2}$, Gourav Khullar$^{19,20,21}$, Jean-Paul Kneib$^{22}$, Ivo Labb\'e$^{23}$, Floriane Leclercq$^{5}$, Rui Marques-Chaves$^{6}$, Charlotte Mason$^{7}$, Kristen B.W. McQuinn$^{18, 24}$, Julian B. Mu\~noz$^{2,3}$, Priyamvada Natarajan$^{25,26}$, Alberto Saldana-Lopez$^{12}$, Mabel G. Stephenson$^{2,3}$, Maxime Trebitsch$^{27}$, Marta Volonteri$^{1}$, Andrea Weibel$^{6}$, Adi Zitrin$^{28}$}
\affiliation{$^1$Institut d'Astrophysique de Paris, CNRS, Sorbonne Universit\'e, 98bis Boulevard Arago, 75014, Paris, France}
\affiliation{$^{2}$Department of Astronomy, The University of Texas at Austin, 2515 Speedway, Stop C1400, Austin, TX 78712, USA}
\affiliation{$^{3}$Cosmic Frontier Center, The University of Texas at Austin, Austin, TX 78712, USA}
\affiliation{$^4$Department of Physics and Astronomy, Tufts University, 574 Boston Ave., Medford, MA 02155, USA}
\affiliation{$^{5}$Univ Lyon, Univ Lyon1, Ens de Lyon, CNRS, CRAL UMR5574, F-69230, Saint-Genis-Laval, France}
\affiliation{$^{6}$Department of Astronomy, University of Geneva, Chemin Pegasi 51, 1290 Versoix, Switzerland}
\affiliation{$^{7}$Cosmic Dawn Center (DAWN), Niels Bohr Institute, University of Copenhagen, Jagtvej 128, Ko benhavn N, DK-2200, Denmark}
\affiliation{$^{8}$David A. Dunlap Department of Astronomy and Astrophysics, \\ University of Toronto, 50 St. George Street, Toronto, Ontario, M5S 3H4, Canada}
\affiliation{$^{9}$Dunlap Institute for Astronomy and Astrophysics, 50 St. George Street, Toronto, Ontario, M5S 3H4, Canada}
\affiliation{$^{10}$MIT Kavli Institute for Astrophysics and Space Research, 70 Vassar Street, Cambridge, MA 02139, USA}
\affiliation{$^{11}$CNRS, IRAP, 14 Avenue E. Belin, 31400 Toulouse, France}
\affiliation{$^{12}$Department of Astronomy, Oskar Klein Centre, Stockholm University, AlbaNova University Centre, SE-106 91 Stockholm, Sweden}
\affiliation{$^{13}$Centre for Extragalactic Astronomy, Department of Physics, Durham University, South Road, Durham DH1 3LE, UK}
\affiliation{$^{14}$Institute for Computational Cosmology, Department of Physics, Durham University, South Road, Durham DH1 3LE, UK}
\affiliation{$^{15}$Department of Physics and Astronomy and PITT PACC, University of Pittsburgh, Pittsburgh, PA 15260, USA}
\affiliation{$^{16}$Leiden Observatory, Einsteinweg 55, NL-2333 CC Leiden, The Netherlands}
\affiliation{$^{17}$OCA, P.H.C Boulevard de l'Observatoire CS 34229, 06304 Nice Cedex 4, France}
\affiliation{$^{18}$Space Telescope Science Institute, 3700 San Martin Drive, Baltimore, MD, 21218}
\affiliation{$^{19}$Department of Astronomy, University of Washington, Physics-Astronomy Building, Box 351580, Seattle, WA 98195-1700, USA}
\affiliation{$^{20}$Institute for Data-Intensive Research in Astrophysics and Cosmology (DiRAC), University of Washington, Physics-Astronomy Building, Box 351580, Seattle, WA 98195-1700, USA}
\affiliation{$^{21}$eScience Institute, University of Washington, Physics-Astronomy Building, Box 351580, Seattle, WA 98195-1700, USA}
\affiliation{$^{22}$Laboratoire d’astrophysique, Ecole Polytechnique F\'ed\'erale de Lausanne (EPFL) Observatoire, CH-1290 Versoix, Switzerland}
\affiliation{$^{23}$Centre for Astrophysics and Supercomputing, Swinburne University of Technology, Melbourne, VIC 3122, Australia}
\affiliation{$^{24}$Department of Physics and Astronomy, Rutgers University, Piscataway, NJ 08854, USA}
\affiliation{$^{25}$Department of Astronomy, Yale University, PO. Box 208101, New Haven, CT 06520-8101}
\affiliation{$^{26}$Department of Physics, Yale University, P.O. Box 208102, New Haven, CT 06520-8102}
\affiliation{$^{27}$LUX, Observatoire de Paris, Universit\'e PSL, Sorbonne Universit\'e, CNRS, 75014 Paris, France}
\affiliation{$^{28}$Department of Physics, Ben-Gurion University of the Negev, P.O. Box 653, Be'er-Sheva 84105, Israel}

\thanks{$^*$E-mail: \href{mailto:hakim.atek@iap.fr}{hakim.atek@iap.fr}}

\begin{abstract}
We present an overview of the \jwst\ GLIMPSE program, highlighting its survey design, primary science goals, gravitational lensing models, and first results. GLIMPSE provides ultra-deep \jwst/NIRCam imaging across seven broadband filters (F090W, F115W, F150W, F200W, F277W, F356W, F444W) and two medium-band filters (F410M, F480M), with exposure times ranging from 20 to 40 hours per filter. This yields a 5$\sigma$ limiting magnitude of 30.9 AB (measured in a $0\farcs2$ diameter aperture). The field is supported by extensive ancillary data, including deep HST imaging from the Hubble Frontier Fields program, VLT/MUSE spectroscopy, and deep \jwst/NIRSpec medium-resolution multi-object spectroscopy. Exploiting the strong gravitational lensing of the galaxy cluster Abell S1063, GLIMPSE probes intrinsic depths beyond 33 AB magnitudes and covers an effective source-plane area of approximately 4.4 arcmin$^2$ at $z \sim 6$. The program's central aim is to constrain the abundance of the faintest galaxies from $z \sim 6$ up to the highest redshifts, providing crucial benchmarks for galaxy formation models, which have so far been tested primarily on relatively bright systems. We present an initial sample of $\sim 540$ galaxy candidates identified at $6 < z < 16$, with intrinsic UV magnitudes spanning \muv\ = $-$20 to $-$12. This enables unprecedented constraints on the extreme faint end of the UV luminosity function at these epochs. In addition, GLIMPSE opens new windows for spatially resolved studies of star clusters in early galaxies and the detection and characterization of faint high-$z$ active galactic nuclei. This paper accompanies the first public data release, which includes reduced \jwst\ and \hst\ mosaics, photometric catalogs, and gravitational lensing models.  
\vspace{0cm}
\end{abstract}

\section{Introduction} 
\label{sec:intro}

The \jwst\ \citep{gardner06,rigby23} has revolutionized the study of the distant Universe. With an unprecedented sensitivity and a wavelength coverage that extends to 30 $\mu$m, \jwst\ provides superior power for peering deeper and further into Cosmic Dawn. After three years of operations, \jwst\ observations have produced some unexpected results \citep[e.g.,][]{naidu22,finkelstein24,harikane23,matthee23b,labbe23b,witstok25}, shedding a new light on the formation of early galaxies and supermassive black holes. 

Wide-field survey programs like CEERS \citep{finkelstein23}, COSMOS-Web \citep{casey23}, and PRIMER \citep{donnan24} are well suited for discovering bright and rare sources from the epoch of reionization through Cosmic Dawn. These surveys have helped reveal an over-abundance of UV-bright galaxies at redshifts $z >10$, compared to theoretical predictions \citep[][]{bouwens23,austin23,harikane24}, as well as ultra-massive galaxies at $z > 10$ that suggest unusually high baryon fractions \citep[][]{casey24}.

 On the other hand, deep surveys, such as NGDEEP \citep[][]{bagley24}, JADES Origins Field \citep[][]{eisenstein23}, cover a small area on the sky, and are better suited to uncover the fainter population of galaxies. The NGDEEP Ultra-Deep Field achieved point-source depths of $m_{5\sigma}\approx30.4$ AB, detecting 38 galaxy candidates at $z\gtrsim9$ down to faint magnitudes. It yields one of the first robust constraints on the faint end at $z\sim9-11$, showing a very steep slope between $\alpha=-2.2$ and $-2.5$, and little evolution between these epochs \citep{leung23}. Using $\sim40$ hours of ultra-deep NIRCam parallel imaging of the MIDIS program \citep{ostlin25}, reaching $m_{5\sigma}\approx30.8$ AB, \citep{perez-gonzalez23} probed galaxies at $8 < z < 13$, targeting the UV LF in the range $-20<$ \muv $< -17$. Based on 44 high-redshift galaxy candidates, the study finds a steep and nearly constant faint-end slope ($\alpha \sim -2.2$) and a modest evolution in galaxy number density and UV luminosity density across this redshift range. 

A complementary approach leverages gravitational lensing by massive galaxy clusters \citep[][]{kneib11}, which magnifies the brightness of intrinsically faint sources  and increases spatial resolution for detailed, small-scale studies of high-redshift galaxies \citep[e.g.][]{vanzella17}. In particular, the Hubble Frontier Fields \citep[HFF;][]{lotz17} program obtained about 840~orbits of \textit{Hubble} Space Telescope and more than 1000~hours of \textit{Spitzer} Space Telescope observations that reached 29~mag in many of the \textit{HST} WFC3/IR bands. These extraordinarily deep images coupled with the natural magnification from gravity by foreground massive clusters revealed intrinsically fainter (by 2 magnitudes on average) and more distant objects than can be detected in blank-field observations. As such, the HFFs contained some of the most distant and faintest galaxies found pre-JWST \citep{livermore17, bouwens17b, atek18,ishigaki18}, which remain unmatched even by the deepest blank-field \jwst\ surveys to date.

\begin{figure}
    \centering
    \includegraphics[width=\linewidth]{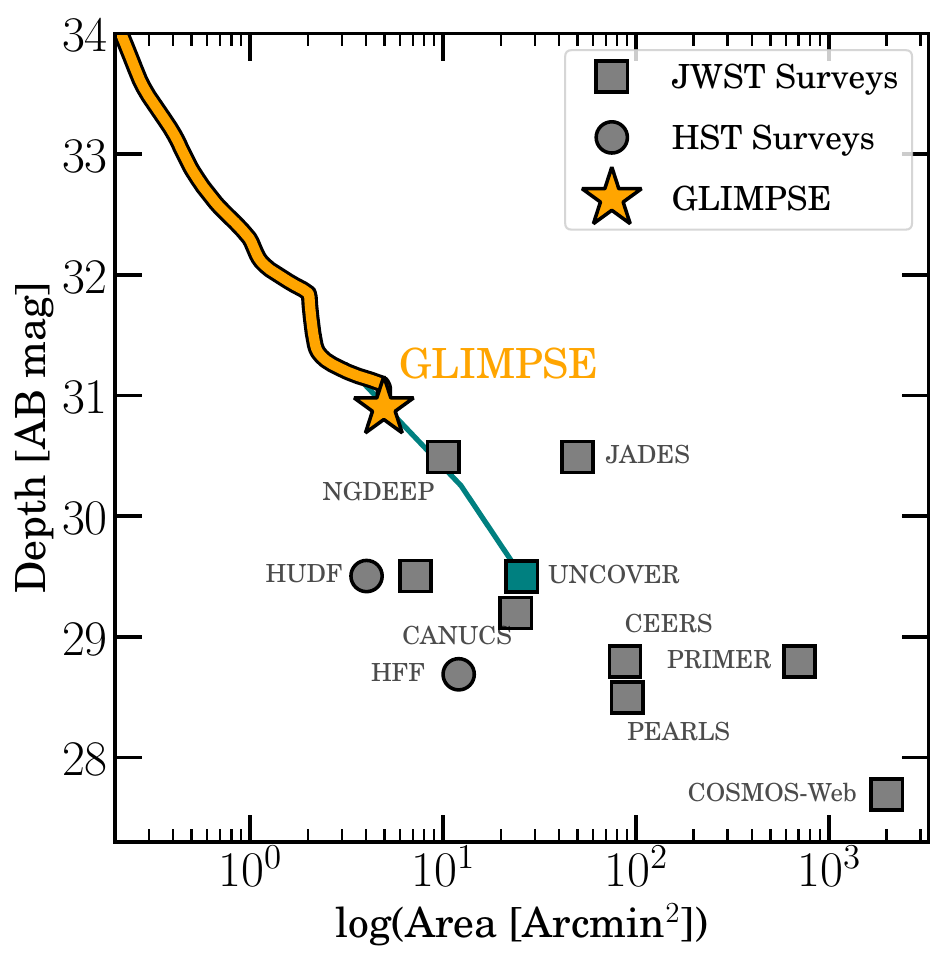}
    \caption{We compare the depth and survey area of various \hst\ and \jwst\ programs with that of GLIMPSE. Wide-field surveys like COSMOS-Web span hundreds of arcmin$^2$ but are relatively shallow, with magnitude limits around 28--29 AB. In contrast, our GLIMPSE survey takes the opposite approach: it probes extremely deep, reaching intrinsic magnitudes fainter than 33 AB, but over a much smaller area in the source (physical) plane. At $z \sim 6$, the total effective survey area of GLIMPSE is 4.5 arcmin$^2$. The orange curve, computed using our lensing model, illustrates how the area is reduced as a function of magnification.}
    \label{fig:depth-area}
\end{figure}

Lensing clusters were key targets in the early \jwst\ observations through the Early Release Science (ERS), Guaranteed Time Observations (GTO), and General Observer (GO) programs. PEARLS and CANUCS are Cycle 1 GTO programs that target lensing clusters. PEARLS (PI Windhorst) combines deep imaging and spectroscopy of seven clusters, While the CANUCS program (PI Willott) targets five lensing clusters and parallel fields using NIRCam, NIRISS, and NIRSpec. Notably, the Abell 2744 cluster system \citep{abell89}, with its large high-magnification area, has been the focus of multiple Cycle 1--3 \jwst\ observations. The ERS program GLASS \citep[PID: ERS 1324][]{treu22} obtained NIRISS, NIRSpec, and NIRCam observations. The NIRCam images consisted of 1.5-6.5~hours across 7 bands (F090W, F115W, F150W, F200W, F277W, F356W, and F444W) reaching 29.7~mag in F444W, the deepest band \citep{treu22}. Subsequent Cycle 1 observations from the Ultradeep NIRSpec and NIRCam ObserVations before the Epoch of Reionization \citep[UNCOVER; PID: 2561][]{bezanson22} provide extraordinarily deep NIRSpec prism spectra \citep{price25} and 4-6~hours in 7 NIRCam filters that reached 29.7~mag at the 5$\sigma$ significance in the F356W band \citep{weaver23}. Further Cycle 2 and 3 programs have added the full suite of medium bands, with the MEGASCIENCE program \citep{suess24}, and even deeper F070W+F090W imaging with the All the Little Things \citep[ALT;][]{naidu24}. These JWST observations of Abell 2744 have pushed the redshift frontier to extraordinary limits through imaging \citep{castellano22, atek23a,chemerynska24a} and spectroscopic confirmation of some of the most distant galaxies \citep{wang23c,fujimoto23b}. These lensed observations have also allowed the characterization of the role of faint galaxies in the process of cosmic reionization \citep{Atek24a}.

In this context, and with the goal of building on the legacy of the {\em Hubble} Frontier Fields in the \jwst\ era, we launched the GLIMPSE\footnote{\url{https://jwst-glimpse.github.io}} program in Cycle 2 (PID: 3293, PIs Atek \& Chisholm). GLIMPSE combines ultra-deep NIRCam imaging with strong gravitational lensing to investigate the formation of the earliest galaxies and uncover the sources driving cosmic reionization. While previous \jwst\ blank-field surveys reach significantly deeper than \hst\ observations within comparable exposure times, they remain limited to galaxies brighter than \muv\ $\approx -17$ (Figure \ref{fig:depth-area}). By exploiting the magnification provided by massive lensing clusters, GLIMPSE extends the accessible UV luminosity function by up to four additional magnitudes, enabling the detection of intrinsically faint galaxies that likely dominate the ionizing photon budget during the epoch of reionization. In this paper, we present an overview of the GLIMPSE survey, its science objectives, and initial results. The paper is organized as follows. In Section~\ref{sec:survey}, we describe the survey design and its main characteristics. Section~\ref{sec:obs} details the \jwst\ observations along with ancillary datasets. The data reduction procedures and photometric measurements are outlined in Section~\ref{sec:reduction}, followed by the strong lensing model presented in Section~\ref{sec:lensing}. In Section~\ref{sec:goals}, we discuss the primary scientific objectives of the program. All magnitudes are reported in the AB system, and we adopt the \textit{Planck} 2018 cosmology \citep{planck18} throughout.

\section{Survey design}
\label{sec:survey}

Given the number of massive galaxy clusters that can serve as powerful gravitational lenses, we carefully select our targets based on several specific criteria, which we discuss here. At the highest magnification, lensing uncertainties dominate both the individual properties of galaxies (such as the luminosity, stellar mass, etc.) and the statistical quantities such as the luminosity function. Because the latter relies on a combination of galaxy luminosity and effective survey volume, it is most impacted by lensing uncertainties. Therefore, our choice of target is primarily motivated by reducing these uncertainties while maximizing the lensing power of the cluster, i.e., the total number of highly-magnified sources. Among the HFF clusters, AS1063 has the most regular morphology, which leads to a simple parametrization of the dark matter distribution. In addition, it is well concentrated to fit the VLT/MUSE field of view, providing more than 70 multiple images with spectroscopic redshifts. In the end, AS1063 has one of the best constrained lensing model \footnote{\url{https://archive.stsci.edu/prepds/frontier/lensmodels/}}. For comparison, Abell 2744 and MACS0717 are morphologically too complex to model, and are extended well beyond the spectroscopic constraints from MUSE and \hst\ imaging, leading to disagreements between modeling teams on the multiple image-systems used in the models, which produce large lensing uncertainties. {This particular disagreement between modeling techniques and teams has been well documented in the HFF lens modeling comparison project \citep{meneghetti17}}

\section{Observations}
\label{sec:obs}

\subsection{JWST observations}

We obtained deep NIRCam imaging of the Frontier Fields lensing cluster Abell S1063 ($\alpha=$\ra{22}{48}{44.13},~ $\delta=$\decl{-44}{31}{57.50}) at a redshift of $z = 0.348$. The compact morphology and the size of the high-magnification region fit well within a single NIRCam module. We selected MODULE B to observe the cluster center, as it offers slightly better sensitivity than MODULE A. Consequently, MODULE A is directed at a ``blank'' region, which is still modestly magnified by the lensing effect of AS1063. For this reason, the lensing model construction includes both modules.

To achieve the desired depth, we require a background level within the 30th percentile of the maximum. To further improve depth and mitigate cosmic ray impact, we used the MEDIUM8 readout mode with 7 to 8 groups per integration. A compact INTRAMODULEBOX dither pattern with 6 positions was selected to cover the short-wavelength intra-module gaps while maximizing the full-depth area. Additionally, we used a subpixel dither with 4 positions to optimize Point Spread Function (PSF) sampling. Following STScI recommendations, we also applied position angle restrictions to avoid detector artifacts (\lq{}\lq{}Claws\rq{}\rq{}) caused by bright stars outside the field of view. A dedicated tool is now provided by STScI\footnote{\url{https://github.com/spacetelescope/jwst_rogue_path_tool}}
 
\jwst\ observations consist of a total of 120 hours of exposure across seven broadband filters and two medium-band filters. Table \ref{tab:depth} provides a breakdown of the exposure times and associated $5\sigma$ depths for each filter. The GLIMPSE survey achieved the deepest imaging in F090W and F115W (39.1 hours each), enabling us to identify the Lyman break of galaxies during the reionization epoch ($z=6-9$). We also acquired longer-wavelength imaging in the F150W/F277W pair (22.3 hours) and the F200W/F356W pair (19.5 hours). This filter combination is crucial for selecting Dark Ages galaxies ($z>12$), whose Lyman break is redshifted to F150W and redder bands. While the F444W exposure was obtained simultaneously with F115W, we split the long-wavelength (LW) channel exposure for F090W into two medium-band exposures in F410M (16.7 hours) and F480M (22.3 hours). These medium-band filters, by encompassing rest-frame optical emission lines, significantly improve redshift determination and constraints on galaxy properties. The emission-line contribution also enables shorter integration times compared to broadband filters. The observing sequence was organized to start with the shortest wavelengths in each visit to mitigate persistence effects.

\begin{figure*}
    \centering
    \includegraphics[width=\linewidth]{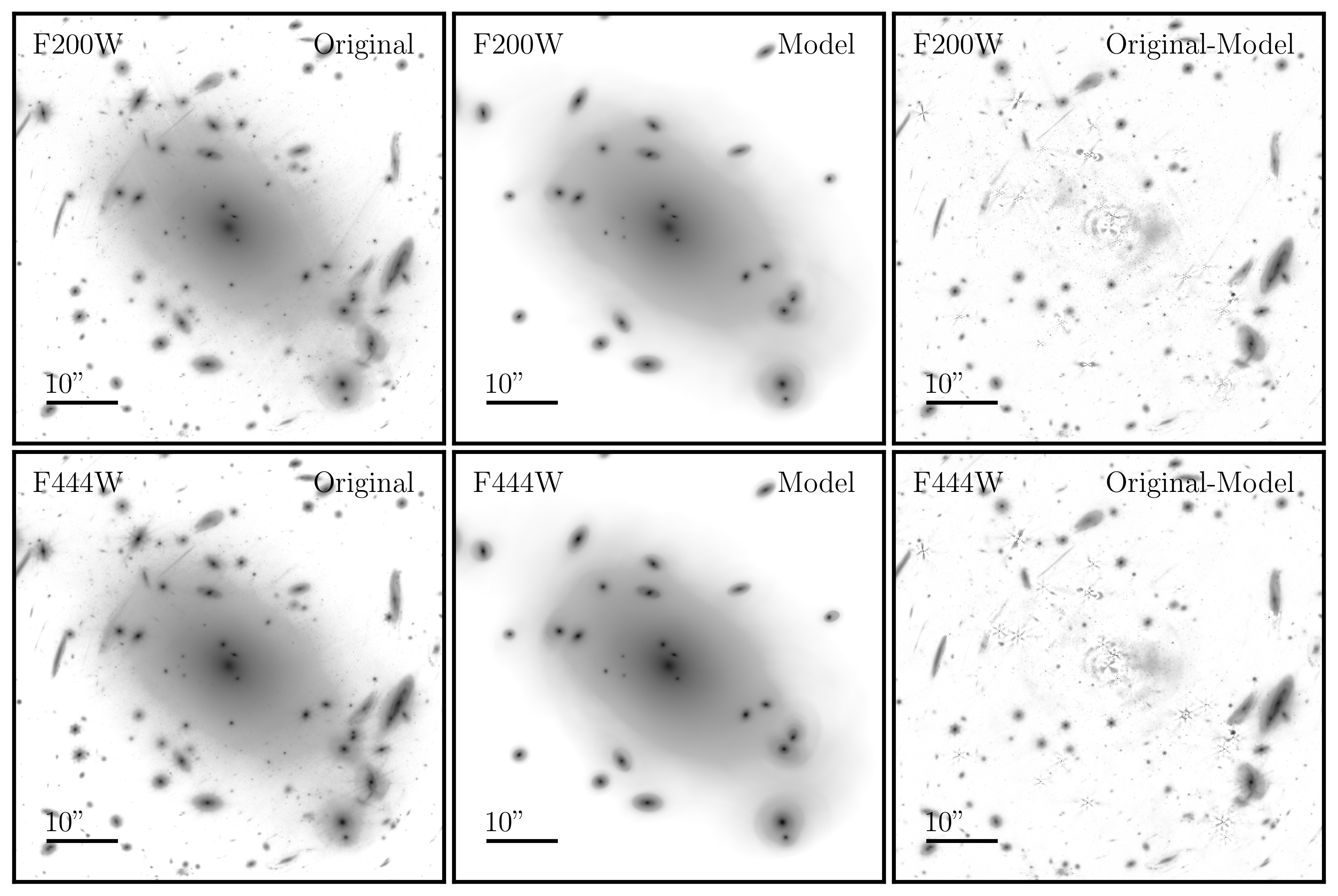}
    \caption{The ICL and bCGs modeling and subtraction. Examples in the SW filter F200W (top row) and LW filter F444W (bottom row) are shown. For each example we show the original image, the fitted foreground model, and the final corrected image.}
    \label{fig:bcg}
\end{figure*}

\subsection{Ancillary data}

In addition to the GLIMPSE observations, our dataset and photometric measurements include supplementary data from both \jwst\ and \hst. Program ID 1840 (PI: Alvarez-Marquez) acquired NIRCam observations of AS1063 in SUBARRAY mode using Module B. The data include SW filters F115W, F150W, F200W, and LW filters F250M, F300M, and F444W. These observations are considerably shallower than GLIMPSE, reaching a $5\sigma$ depth of approximately 28.5--29 mag.

This lensing cluster has been a focus of extensive \hst\ observations as part of the Hubble Frontier Fields (HFF) program (PID 14037, PI Lotz). Optical imaging was collected using the Advanced Camera for Surveys (ACS) in three broad-band filters: F435W, F606W, and F814W. Near-infrared (NIR) imaging was obtained with the Wide Field Camera 3 (WFC3) in four filters: F105W, F125W, F140W, and F160W. ACS and WFC3 imaging was conducted in parallel, covering both the cluster core and an adjacent flanking field. These observations reach a $5\sigma$ limiting magnitude of approximately 29. The BUFFALO (PID 15117, PI Steinhardt) program further extended imaging coverage of both the cluster and the parallel field. The resulting observations comprise a 2$\times$2 mosaic of ACS optical data in the F606W and F814W filters, as well as WFC3/IR data in the F105W, F125W, and F160W filters, reaching a $5\sigma$ depth of approximately 27.5 magnitudes. Additionally, we incorporate deep optical imaging data in two long-pass filters F200LP and F350LP from the FLASHLIGHTS program (PID 15936). Spectroscopic observations of AS1063 were conducted with the VLT/MUSE instrument in two separate pointings (program IDs 60.A-9345 and 095.A-0653), and more recently as part of the GLIMPSE follow-up programs (PIDs 114.28L1.001 and 116.28VK.001, PIs Richard and Fujimoto). These data are instrumental in building the strong lensing model. In addition, a $\sim$ 50h ALMA program (PID 2024.1.00876.S, PI Kokorov) covers the 2\arcsec $\times$2\arcsec cluster center with continuous 30 GHz spectral scans. X-ray observations (123 ks) of the AS1063 field (16\arcsec$\times$16\arcsec) were obtained by {\em Chandra} as part of multiple programs (IDs 4966, 18611, 18818).

\section{Data Reduction and Photometry}
\label{sec:reduction}

The data products of this first release can be found on the GLIMPSE website \footnote{\url{https://jwst-glimpse.github.io}} and the public archive \footnote{\url{https://doi.org/10.5281/zenodo.22552278}}. 

\subsection{Data reduction}
The reduction of the NIRCam imaging data begins with processing the individual uncalibrated {\tt *rate.fits} files using the official \jwst\ Science Calibration Pipeline (version {\tt v12.0.9}), in conjunction with the Calibration Reference Data System (CRDS) context file {\tt jwst\_1321.pmap}. In addition to the standard procedures provided by STScI, we apply several custom processing steps to the output {\tt *uncal.fits} files aimed at mitigating various artifacts. These include the removal of ``snowballs'' and ``wisps'' correction of cosmic ray hits and persistence following \citet{endsley23}, and suppression of diffraction spikes originating from bright stars outside the field of view \citep{rigby23,bradley23,bagley23}. 

A key component of our custom reduction procedure involved generating improved flat-field calibration frames using all publicly available NIRCam imaging data as of January 12, 2025. The use of standard STScI flat fields introduced significant correlated noise in the final images, reducing the effective depth and leading to the appearance of numerous spurious sources. This issue is particularly relevant given the GLIMPSE observing strategy, which employs six large dithers combined with four subpixel dithers at each position to enhance PSF sampling. While effective for PSF reconstruction, this strategy also leads to the co-addition of flat-field imperfections, which can manifest as high apparent signal-to-noise ratio sources, especially in regions near the edges of the final mosaic where fewer large dithers overlap. This procedure leads to an improvement in the imaging depth by up to 0.5 magnitudes in the LW filters and up to 0.2 magnitudes in the SW filters, respectively.  

In the next step, the corrected {\tt *uncal.fits} files are processed through Stage 2 of the JWST Science Calibration Pipeline. The resulting {\tt *cal.fits} files are further corrected for residual 1/\emph{f} noise by applying sigma-clipping along image rows and columns, after masking sources and artifacts \citep{schlawin20}. Following this, we perform a two-dimensional background subtraction using the {\tt sep} package \citep{barbary16}, following the methodology described in \citet{bagley23}.

Finally, all {\tt *cal.fits} frames are processed through stage 3 of the calibration pipeline to create the final mosaics. The output pixel scale is 20 mas~pix$^{-1}$ for SW bands and 40 mas~pix$^{-1}$ for the LW bands, respectively. The mosaics were astrometrically aligned to 3 stars that were selected from GAIA. Our final images reach a $5\sigma$ aperture-corrected depths of 30.8--30.9 mag in D=0\farcs{2} apertures across all broad-bands. 

At the same time, we reduced the ancillary \hst\ ACS and WFC3 images, which are also aligned to Gaia and include data from the following programs:  the Hubble Frontier Fields \citep{lotz17}, BUFFALO \citep{steinhardt20}, and FLASHLIGHTS. HST mosaics were retrieved from the CHArGE archive \citep{kokorev22}, which is hosted on the Dawn JWST Archive \citep{valentino23}. The final output mosaics were drizzled to a pixel scale of 40mas pix$^{-1}$.

\subsection{Modeling and subtracting foreground structures}

To achieve our science goals, it is critical to first mitigate contamination from the foreground light of numerous bright cluster galaxies (bCGs) and the diffuse intra-cluster light (ICL). The depth of our survey magnifies the impact of this contamination, prompting further refinement in our methods for removing bCG and ICL light. Such processing is essential to ensure reliable photometry of background sources, particularly rare high-$z$ galaxies whose light may otherwise be distorted or even rendered undetectable.

We adopt the catalog of bCGs in Abell 1063 from \citet{shipley18}, which is based on the method described by \citet{ferrarese06}. We do not include additional bCGs located outside the original HFF footprint, as their contribution to foreground contamination is negligible. To reduce computational load and data storage, we create a ``cropped mosaic'' to only include a smaller region that encompasses all relevant sources.

Following the methods in \citet{ferrarese06, shipley18, weaver24a}, we begin by generating a mask for all sources not selected for modeling. Source detection is performed using \textsc{SExtractor} \citep{bertin96} with the following parameters: \texttt{DETECT\_THRESH} = 1.2, \texttt{DEBLEND\_NTHRESH} = 10, \texttt{DEBLEND\_MINCONT} = 0.01. This initial pass identifies isolated foreground and background sources. We then run a second detection on the masked image, which excludes the isolated sources, allowing us to identify the remaining structures, typically the cluster members to be modeled and subtracted. Finally, we smooth this mask with a Gaussian kernel to better capture the extended emission from the diffuse and complex ICL.

\begin{figure*}
    \centering
    \includegraphics[width=\linewidth]{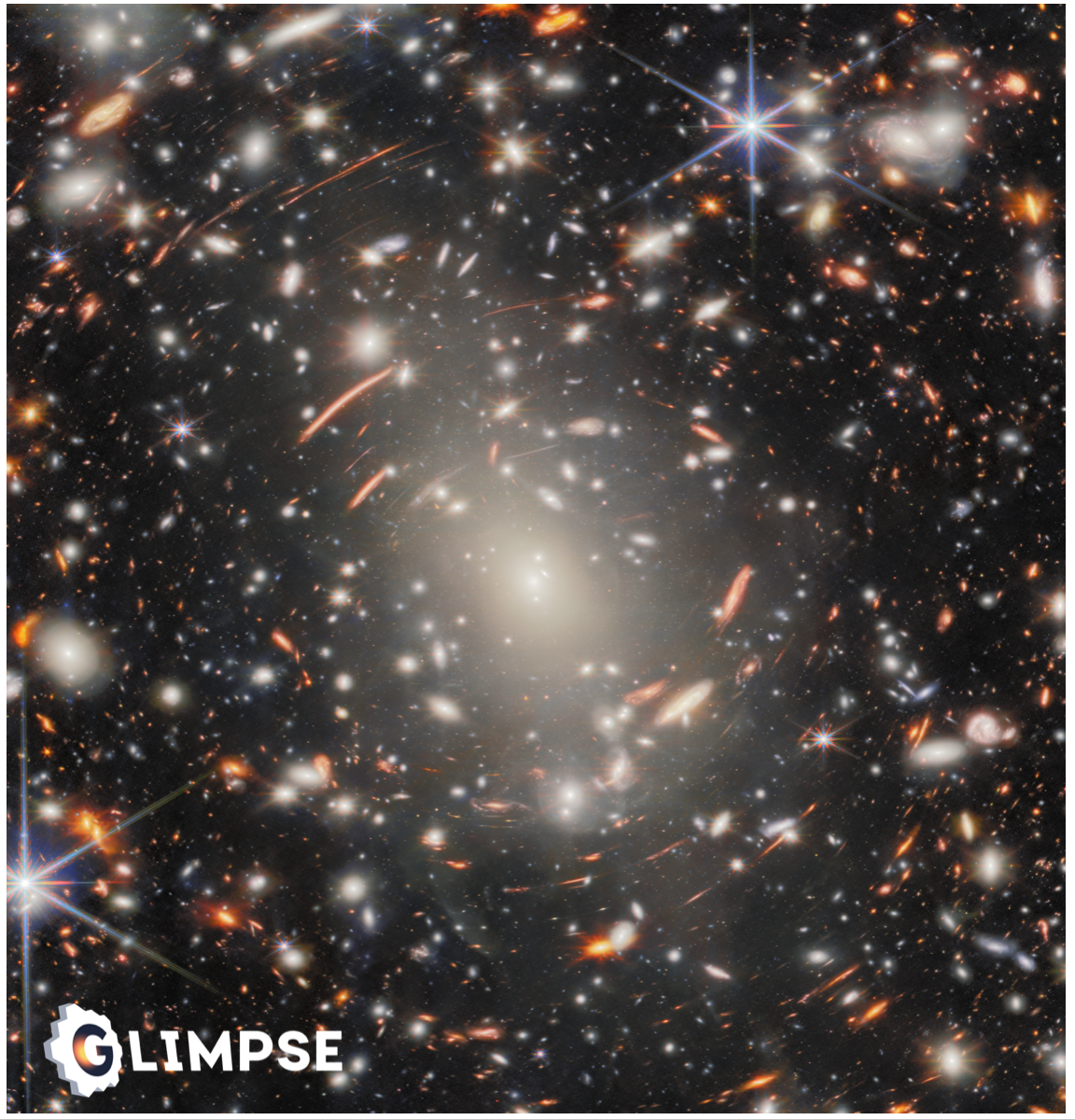}
    \caption{Composite color image of the cluster center from the NIRCam imaging of Module B, which represents the intrinsically deepest image obtained by \jwst. Numerous giant arcs and lensing features are visible, extending from around the BCG and to the outskirts. Compact and point-like sources are seen across all the image, and include globular clusters within the galaxy cluster and distant ultra-faint galaxies. Many distant galaxies are also spatially resolved, allowing the study of individual star clusters. The image is a combination of all 7 broadband and 2 medium-band filters\footnote{\url{https://esawebb.org/images/potm2505a/}}}
    \label{fig:color-image}
\end{figure*}

       

\begin{table}
    \centering
    \begin{tabular}{c|c|c|c}
        Filter & Exposure time [h] & $5-\sigma$ depth & PSF FWHM [mas] \\
        \hline
        \hline
        \multicolumn{4}{c}{JWST} \\
        \hline
        F090W  & 39.1 & 30.85 & 63 \\
        F115W  & 39.1 & 30.87 & 63 \\
        F150W  & 22.3 & 30.77 & 67 \\
        F200W  & 19.5 & 30.82 & 78 \\
        F277W  & 22.3 & 30.82 & 130 \\
        F356W  & 19.5 & 30.77 & 146 \\ 
        F410M  & 16.7 & 30.07 & 170 \\
        F444W  & 39.1 & 30.68 & 177 \\ 
        F480M  & 22.3 & 29.24 & 195 \\
        \hline
        \multicolumn{4}{c}{HST} \\
        \hline
        F200LP & 13.6 & 31.06 & 172 \\
        F350LP & 13.7 & 30.72 & 171 \\
        F390W  & 7.4  & 28.31 & 168 \\
        F435W  & 32.6 & 30.28 & 122 \\
        F475W  & 6.4  & 28.67 & 146 \\
        F606W  & 37.3 & 30.34 & 121 \\
        F814W  & 85.8 & 30.37 & 118 \\
        F105W  & 43.7 & 29.96 & 193 \\
        F125W  & 24.6 & 29.53 & 207 \\
        F140W  & 20.3 & 29.57 & 205 \\
        F160W  & 46.7 & 29.47 & 211 \\
    \end{tabular}
     \caption{Characteristics of the observations in each JWST/NIRCam and HST filter. The $5-\sigma$ limiting magnitudes are calculated in D=0\farcs{2} apertures. The empirical Point Spread Function (PSF) Full Width at Half Maximum (FWHM) measurements are also provided.}
    \label{tab:depth}
\end{table}

To create our initial model, we first isolate each individual cluster member to be modeled by masking out all other adjacent sources. We then use \texttt{ELLIPSE} from the \texttt{ IRAF/Isophote} package to measure the isophotal parameters that are then passed to \texttt{BMODEL} to generate the galaxy model. This initial galaxy model is subtracted from the cropped mosaic. We then rerun this process for each cluster member, which produces our first residual image. To first order, these models and residuals provide a good approximation of the background field. However, we adopt an iterative approach to improve our models and our residuals as done in \citet{shipley18, weaver24}.

Oftentimes, the main challenge in modeling these cluster members is the contaminating light from other nearby bCGs. To address this, we adopt an iterative approach. We first fit the galaxy, then we re-model each galaxy by subtracting the updated best-fit model, and repeating this process for all other members. This yields improved models that more accurately represent the intrinsic light profiles of individual galaxies, rather than blended profiles that are influenced by nearby sources. We repeat this iterative cycle 10 times and average the resulting images to obtain the final model.

During the averaging procedure, we reject the lowest 4 and highest 2 values on a pixel-by-pixel basis. Following the method described in section 3.1.3 of \citet{shipley18}, we use the IRAF task \texttt{IMCOMBINE} with the following parameters: \texttt{combine} = ``average'', \texttt{reject} = ``minmax'', \texttt{nlow} = ``4'', and \texttt{nhigh} = ``2''. Lastly we subtract the averaged best-fit galaxy model from the original cropped mosaic to produce our final averaged residual cropped mosaic. For crowded regions, we re-run \textsc{SExtractor} to generate an updated mask based on the new residual image. This refinement significantly improves the modeling of bCGs that are tightly clustered, particularly those located near the cluster cores.

\begin{figure*}
    \centering
\includegraphics[width=0.8\textwidth]{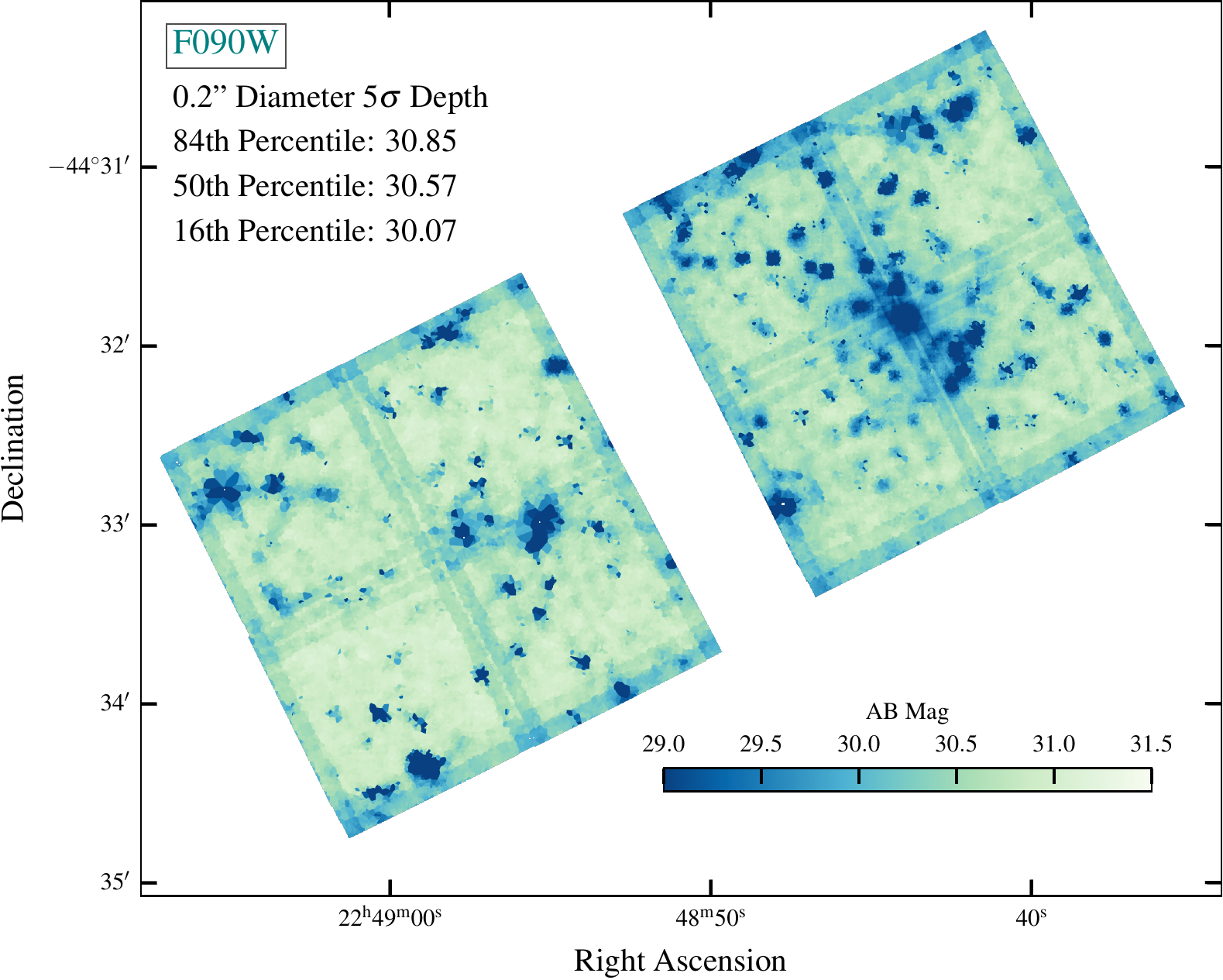}
    \caption{Depth map in the F090W band for the two NIRCam modules. The 5$\sigma$ magnitude limits are derived from flux measurements in randomly placed apertures of diameter $0\farcs{2}$ across the image. The 16th, 50th, and 84th percentile values are reported.}
\label{fig:depth}
\end{figure*}

\begin{figure*}
    \centering
\includegraphics[width=0.8\textwidth]{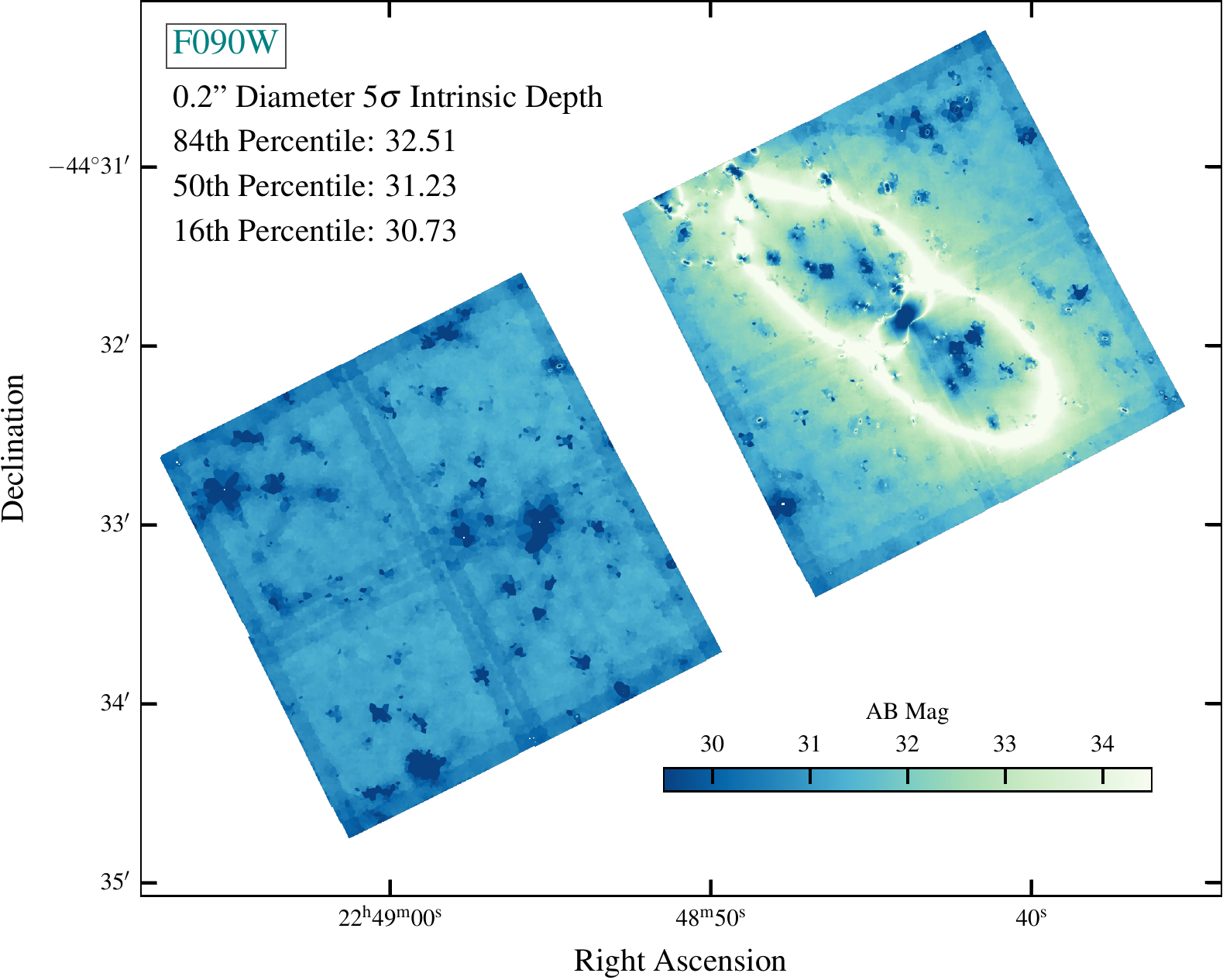}
    \caption{Same as Figure \ref{fig:depth}, but for the intrinsic depth corrected for lensing using the magnification map.}
    \label{fig:depth_intrinsic}
\end{figure*}

We find that averaging over multiple iterations helps suppress noise and reduces variance introduced in any single model run. This approach also mitigates potential biases toward any one galaxy model, which is particularly important given the difficulty of recovering the true light profile of individual bCGs, often blended with light from neighboring cluster members. We show the results of this process and how it effectively removes bCG and ICL contamination near the cluster core in Figure~\ref{fig:bcg}. Some residuals are still present in the final image due to the complex structure of the ICL, as it can be seen in the south-west region of the cluster core. This usually leads to higher uncertainties in the photometric measurements, which rely on apertures around the source for background estimates (see Section \ref{subsec:photometry})

Lastly, we insert our final average residual cropped mosaic back into the original mosaic to undo the initial crop. We perform another background subtraction on this mosaic to smooth any over- or under-subtraction near the edges of the galaxy models. We use the \textsc{SExtractor} \texttt{AUTO} setting with the same parameters as \citet{weaver24a}: mesh size of 192 for SW bands (20 mas pix$^{-1}$) and 96 for LW bands (40 mas pix$^{-1}$), limiting magnitude of 15, and a maximum threshold of 0.01. Because of the highly clustered nature of the central core, we perform another background subtraction that more aggressively removes any contamination. 

Throughout this process, we never introduce the PSF models into our bCG modeling process. This is because the procedure is computationally expensive, and we are not interested in the morphology of the bCGs themselves. As such, the residual image will leave PSF diffraction spikes that will in turn be detected as potential sources. To remove such sources, we mask objects detected around the bCGs as described in the flagging process of \ref{subsec:photometry}. 

The official GLIMPSE color image is shown in Figure \ref{fig:color-image}. Figure \ref{fig:depth} shows the observed depth map, while Figure \ref{fig:depth_intrinsic} shows the intrinsic depth when accounting for lensing magnification.

\begin{figure*}
    \centering
    \includegraphics[width=\linewidth]{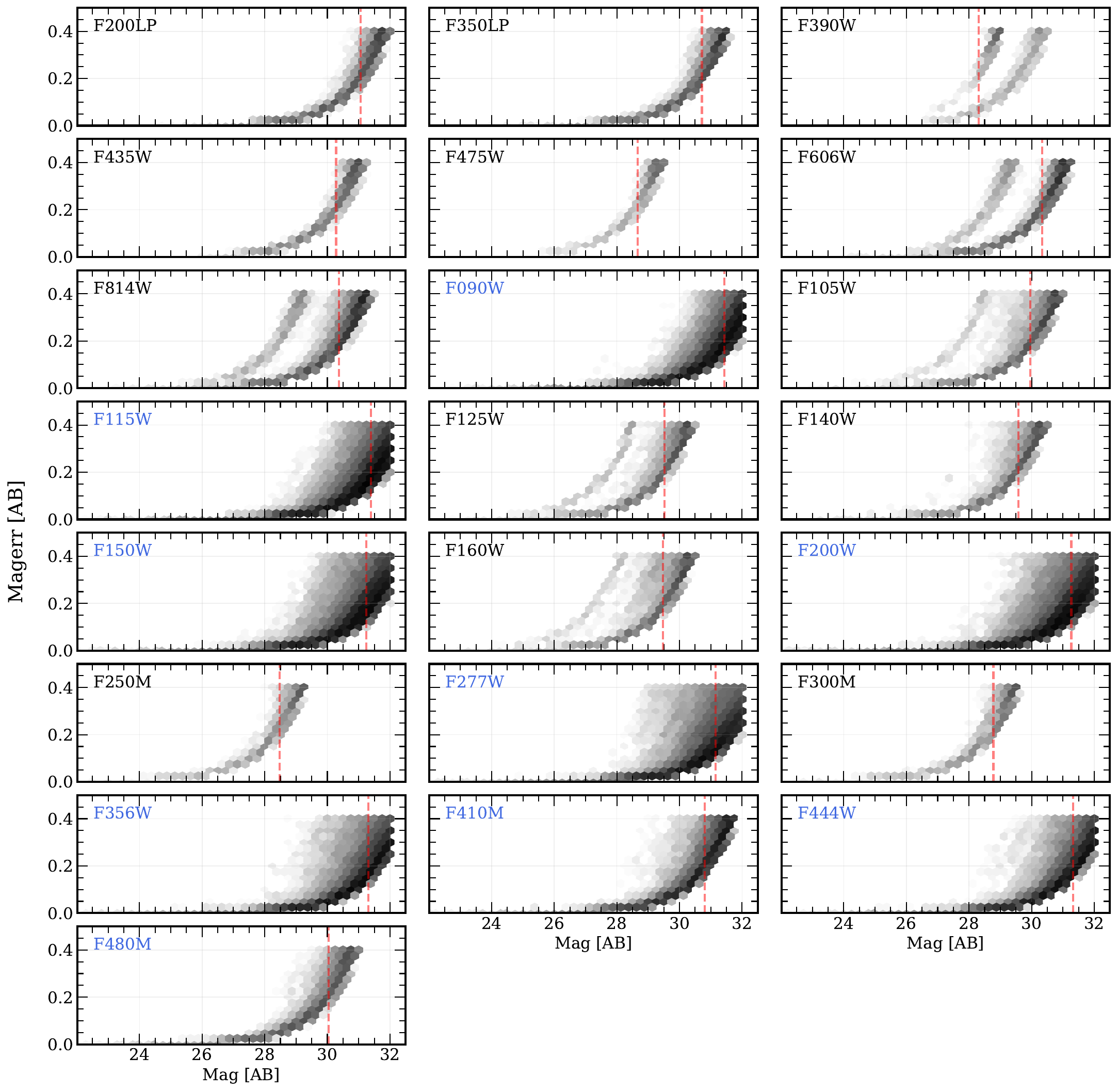}
    \caption{2-D Density plots showing the photometric uncertainty as a function of the source magnitude, measured in D=0\farcs{2} apertures. The GLIMPSE filters are marked in blue, while ancillary data filters are marked in gray. The split between some \hst\ filter curves (in particular F606W and F814W) arises from non-uniform exposure depths across the observed field. The vertical dashed line denotes the $5-\sigma$ depth in each band.}
    \label{fig:magerr}
\end{figure*}

\subsection{Photometry}\label{subsec:photometry}
Prior to extracting photometry, we have homogenized all our \textit{HST} and \textit{JWST} images to our lowest-resolution PSF F480M. This is done to ensure that the color information extracted across PSFs of varying size is consistent across all bands. We start this procedure by constructing a PSF for each band from all the available non-saturated stars in the field, and then applying the convolution kernels to the bCG subtracted mosaics.

Source detection is performed using \textsc{SExtractor} \citep{bertin96}. We first construct inverse-variance-weighted (IVW) signal-to-noise stacks of the non-PSF-convolved short-wavelength (SW) bands (F090W, F115W, F150W, and F200W), from which we generate an initial detection catalog and segmentation map. While detection in the SW bands offers a factor of $\sim$2 improvement in spatial resolution compared to the long-wavelength (LW) bands, it may miss very dusty or high-redshift sources that are faint or undetected in the SW regime. To identify such sources, we also construct an IVW S/N stack of the broad LW bands (F277W, F356W, F444W) and apply the same detection procedure.

Finally, the SW and LW detection catalogs are then merged as follows: we retain all sources from the SW catalog and include only those LW-detected sources whose central pixel falls in a region of the SW segmentation map with zero value, indicating no previously associated source. Our final catalog contains 73,621 sources.

Photometry is performed using \textsc{PHOTUTILS} \citep{bradley20}, with flux densities measured in circular apertures of varying diameters, ranging from $D = 0\farcs{1}$ to $1\farcs{4}$. Photometric uncertainties are computed independently for each object and filter, accounting for both aperture size and spatial variations in image depth. To estimate the background noise, we place 2000 circular apertures in source-free regions near each object, as defined by the segmentation map. The standard deviation of the fluxes within these empty apertures, combined in quadrature with the Poisson noise contribution, defines the final uncertainty. This empirical approach provides a more accurate estimate than weight or error maps alone, as it captures the impact of correlated noise \citep[e.g.,][]{endsley23,weaver23}. Figure \ref{fig:magerr} shows the photometric uncertainties as a function of magnitude for each \jwst\ and \hst\ band used in our catalog. Aperture corrections are applied assuming a point-source profile. We adopt the empirical, symmetric F480M PSF curve of growth to determine the fraction of total flux outside each aperture and apply this as a multiplicative correction factor. We note that we adopted the fixed-aperture size approach instead of the {\tt Kron} \citep{kron80} elliptical apertures because the latter is severely impacted by crowding, source blending, and ICL contamination in this cluster field. As a result, fixed circular apertures are more stable and better suited for color measurements across bands, which is a central point in most of the science enabled by \glimpse\ data. However, we stress that for very extended sources, the measured flux can be underestimated by the aperture-correction photometry. In addition to size parameters, the public \glimpse\ catalog contains the {\tt Kron radius} that allows one to recover the total flux in elliptical apertures.

Finally, to provide the best quality of our photometric catalogs we construct a series of flags, which aim to remove objects whose photometry might be contaminated by artifacts, insufficient depth, or objects that are adjacent to  bCGs. \texttt{flag\_edge} denotes objects for which the x,y centroid position falls within 15 pixels from the edge of the LW detection image. This region near the edges contains fewer overlapping dithers which allows for flat-fielding uncertainties to dominate. \texttt{flag\_spike} contains sources that fall within the manually constructed diffraction spike mask, and are also oriented in the same direction as the spike. \texttt{flag\_bcg} contains all the objects that fall within a $D$=0\farcs{3} circular aperture that intersect an $R$=2\farcs{0} aperture centered on each bCG. Subtraction residuals are often present after removing bCGs which can artificially inject spurious sources near the bCGs.   In each case, =1 signifies that a source satisfies the criterion, and =0 signifies that the source does not satisfy the criterion. Together, we build a single \texttt{use\_phot} flag which removes undesirable objects from all the above categories, where =1 is not flagged for any reason and =0 if any of the above flags are triggered. In other words, a \texttt{use\_phot}$=1$ means that the photometry can be used without severe contamination.
In total, our catalog contains 64,828 sources in the single GLIMPSE pointing for which \texttt{use\_phot}=1. The photometric redshifts derived using {\sc Eazy} range from $z\sim0$ to $z\sim30$.

\section{Strong Lensing model}
\label{sec:lensing}

\begin{figure*}
    \centering
    \includegraphics[width=\linewidth]{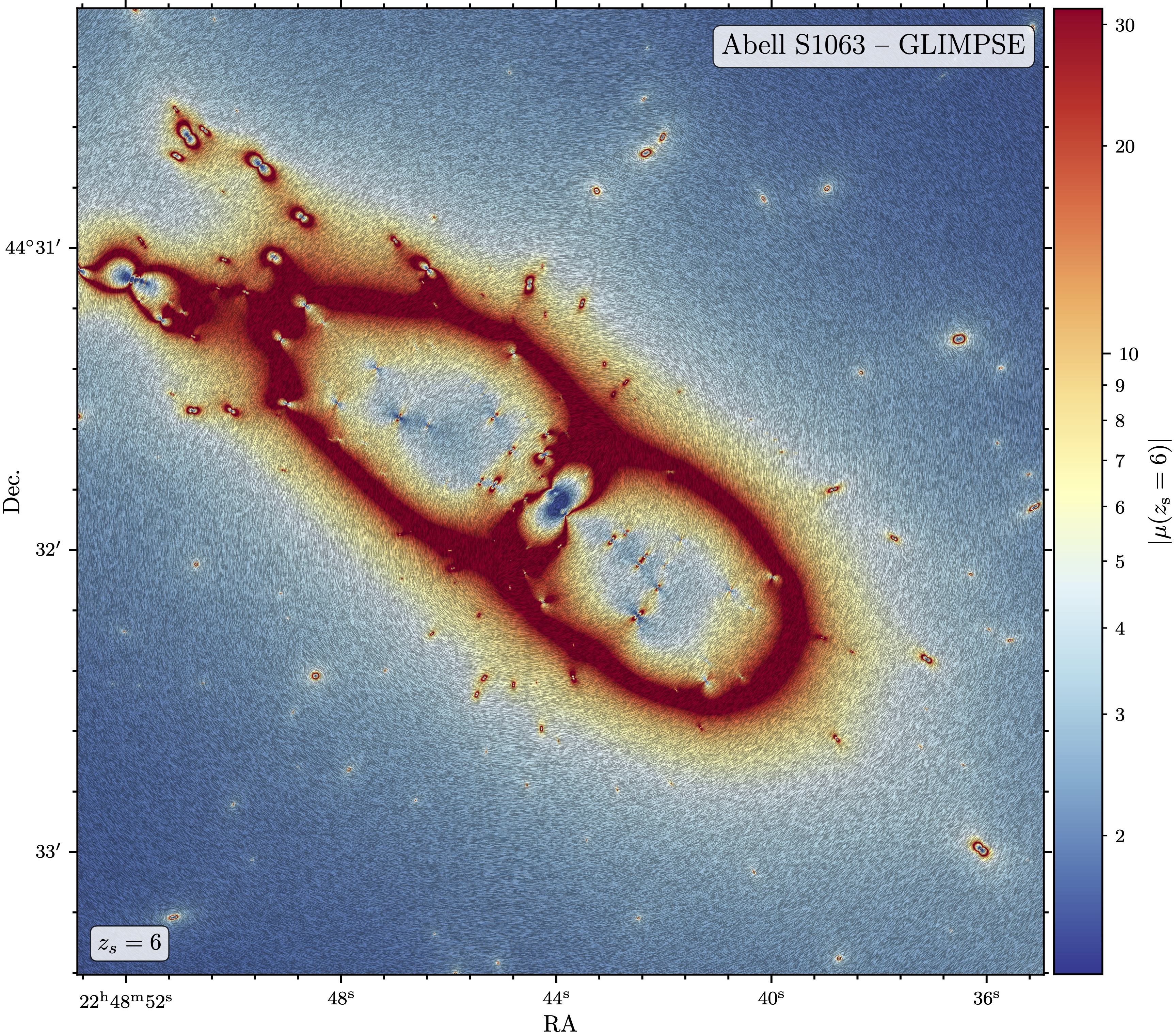}
    \caption{Strong lensing model of the galaxy cluster Abell S1063. The magnification map is color-coded by the gravitational magnification factor, ranging from $\mu \sim 2$ in the blue regions to $\mu > 100$ in the red regions near the critical curve for a source at $z = 6$. Overlaid flow lines indicate the direction of the lensing shear, illustrating how background galaxies are distorted by the cluster's gravitational potential.}
    \label{fig:lensing}
\end{figure*}

As one of the \textit{The Cluster Lensing and Supernova Survey with Hubble} \citep[CLASH;][]{postman12} and \textit{Hubble Frontier Fields} \citep[HFF;][]{lotz17} clusters, AS1063 has been extensively studied and modeled with its strong gravitational lensing (SL) effect \citep[e.g.][]{monna14,zitrin2015,caminha16,bergamini2019,beauchesne2024}. Building on these works, we construct a new parametric SL mass model of AS1063 for GLIMPSE using an updated version of the parametric method by \citet{zitrin2015}, which has been revised to be fully analytic and thus not limited to a grid resolution. This code, sometimes referred to as \texttt{Zitrin-analytic} in recent works, has already been successfully deployed on several SL clusters with JWST \citep[e.g.][]{pascale2022,furtak23b}.

Similarly to all parametric lensing methods, we model the cluster with two main components: A smooth dark matter (DM) component, parametrized with pseudo-isothermal elliptical mass distributions (PIEMDs) \citep{kassiola93}, and the cluster member galaxies which are parametrized as dual pseudo-isothermal ellipsoids (dPIE) \citep{eliasdottir07}. We model AS1063 with two cluster-scale DM halos, one centered on the BCG, the other on a group of galaxies to the north-east, following \citet{bergamini2019} and \citet{beauchesne2024}. The model further comprises 303 cluster member galaxies. For SL constraints, we start from the \citet{beauchesne2024} sample of spectroscopically confirmed multiple images in AS1063 and add 6 photometric systems, in particular two systems to also constrain for the first time the north-eastern sub-halo. The model is thus constrained with {a total of 79 multiple images of 30 sources, 24 of which have spectroscopic redshifts, detected with the HST and MUSE data. We optimize the model in the source plane using a set of Monte-Carlo Markov Chains (MCMC), as detailed in e.g.\ \citet{furtak23b}, and achieve a final average lens plane reproduction error of $\Delta_{\mathrm{RMS}}=0.54\arcsec$. This preliminary version of the model is shown in Fig.~\ref{fig:lensing} and has already been used in \citet{kokorev25} and \citet{fujimoto25}.

As has been shown with e.g.\ SMACS~J0723.3-7327 \citep[e.g.][]{pascale2022,mahler23}, Abell~2744 \citep[][]{furtak23b,bergamini23b}, or MACS~J0416.1-2403 \citep[e.g.][]{rihtarsic25}, \jwst/NIRCam observations of SL clusters yield plethora of spectacular and previously unknown multiple-image systems and lensed background objects. Given the unprecedented depth of GLIMPSE, and \jwst's higher resolution and wavelength coverage, we detect 45 new multiple-image systems in AS1063. With these, and with new MUSE (ESO programs 114.28L1.001 and 116.28VK.001) and \jwst/NIRSpec observations from our spectroscopic follow-up program (PID 9223, PIs Fujimoto \& Naidu), we will be able to significantly improve the SL models of AS1063 and perhaps constrain the DM potential anomaly in the south-west discussed in \citet{beauchesne2024}. An updated SL mass model of AS1063, with spectroscopic redshifts from the DDT NIRSpec follow-up, achieves a lens-plane $\Delta_{\mathrm{RMS}}=0.32\arcsec$, and is described in detail in \citet{furtak26}.

\section{Science goals and early results}
\label{sec:goals}

The broad spectral coverage of the GLIMPSE observations enables us to identify and photometrically characterize a diverse range of sources, from early star-forming galaxies to low-mass quiescent systems, across a wide span of redshifts. Representative examples are shown in Figure~\ref{fig:filters}, where we present typical source templates along with their simulated photometry across all bands.

\begin{figure*}
    \centering
    \includegraphics[width=\linewidth]{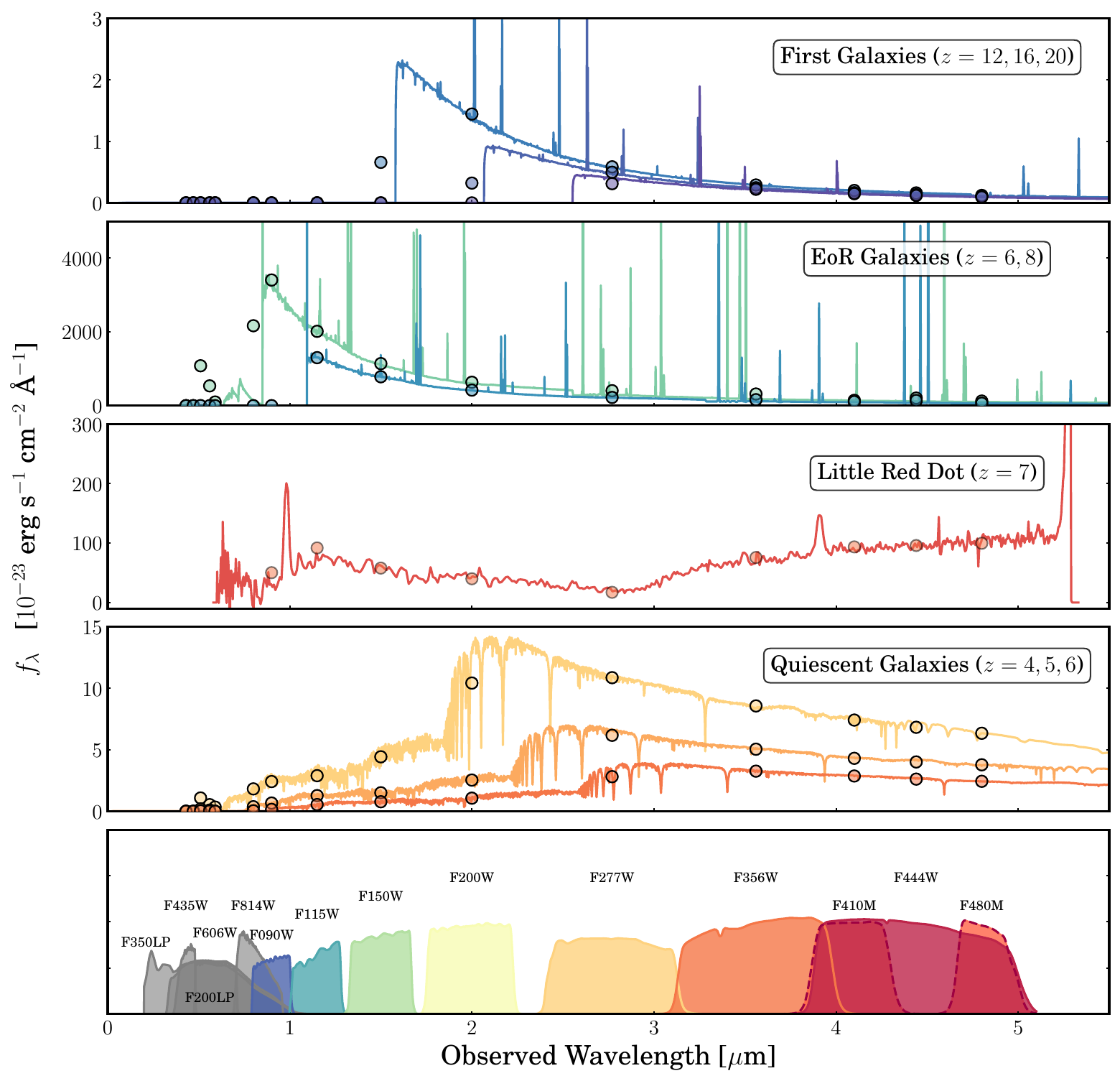}
    \caption{Model SEDs of the typical sources explored by the observing strategy of GLIMPSE together with simulated flux densities measured in each \hst\ and \jwst\ filter. {\em First row} shows early star-forming galaxies at redshifts $z=12$, 16 and 20. {\em Second row} shows young star-forming galaxies at the epoch of reionization ($z=6-8$). {\em Third row} shows a typical Little Red Dot (LRD) spectrum at $z\sim7$ \citep{furtak24b}, with a characteristic V-shape, easily identifiable in photometric data. {\em Fourth row} shows low-mass quiescent galaxies with log(\mstar/\msol)=6,7,8 at redshifts $z=4,5,6$. {\em Last row} summarizes the NIRCam imaging bands used by GLIMPSE (color-shaded) and the ancillary \hst\ bands (gray-shaded).   
    }
    \label{fig:filters}
\end{figure*}

\begin{figure*}
    \centering
    \includegraphics[width=0.85\linewidth]{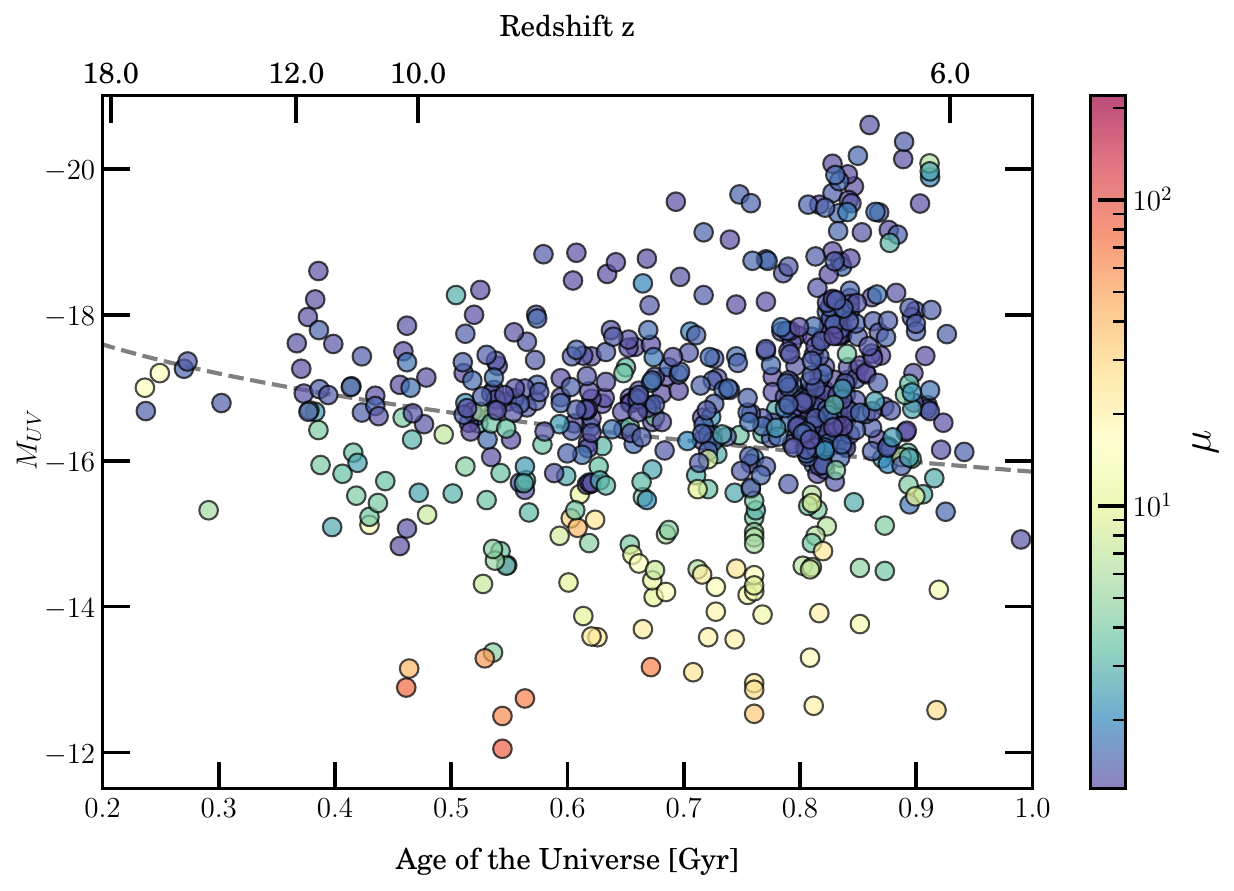}
    \caption{Compilation of high-redshift galaxies identified in the GLIMPSE survey, with their intrinsic absolute UV magnitude as a function of photometric redshift and the corresponding age of the Universe \citep{kokorev25}. Each point is color-coded by the amplification factor $\mu$. The dashed line indicates the \muv\ detection limit as a function of redshift for the \jwst\ Origins Field (JOF) observations \citep{eisenstein23}. Thanks to strong gravitational lensing, GLIMPSE is able to probe significantly fainter galaxies across all redshifts compared to unlensed surveys}
    \vspace{0.3cm}
    \label{fig:muv-z}
\end{figure*}

\subsection{The faint end of the UV luminosity function}

The shape and amplitude of the ultraviolet luminosity function (UV LF) at the faint end, and their evolution across the epochs of Cosmic Dawn ($z>9$) and Cosmic Reionization ($6<z<9$), offer powerful constraints on galaxy formation models. 

Theoretical predictions for the abundance of ultra-faint galaxies with \muv$ > -16$ span several orders of magnitude, reflecting significant uncertainties in star formation efficiency and feedback processes. The main challenges in simulating galaxy populations stem from: (a) the absence of a robust theoretical framework for star formation in molecular and atomic gas clouds, and (b) the complex role of stellar and supernova (SN) feedback in regulating further star formation. These processes remain unresolved in current numerical simulations, leading to the adoption of diverse phenomenological models that must be calibrated against observed galaxy abundances \citep[see][for a review]{naab17}. However, constraining the faint end of the UV LF remains a major challenge. Most JWST surveys to date have primarily targeted the bright end, leaving the key faint regime largely unexplored. The GLIMPSE survey addresses this gap by combining the power of strong gravitational lensing with ultra-deep imaging to access the faintest galaxies at $z > 6$.

Using a combination of Lyman-break selection and photometric redshift estimates derived from spectral energy distribution (SED) fitting with the {\sc Eazy} software \citep{Brammer2008}, we have assembled a sample of nearly 540 galaxy candidates at $z > 6$ (Figure~\ref{fig:muv-z}). Specifically, we identify 411 sources in the redshift range $6 < z < 9$, with absolute UV magnitudes spanning \muv\ = [$-20$, $-12$] \citep{atek26}, 114 sources at $9 < z < 11$, with \muv\ = [$-19$, $-12$], and 11 sources at $11 < z < 15$, with luminosities in the range \muv\ = [$-19$, $-13$] \citep{chemerynska25}. Figures \ref{fig:cutouts1} and \ref{fig:cutouts2} shows examples of galaxy candidates with their best-fit SEDs and photometric redshift probability distribution. Additionally, we identify two galaxy candidates at $z \sim 16$, with absolute magnitudes around \muv\ = $-17$, which are discussed in detail in \citet{kokorev25}. 

Figure \ref{fig:muv-z} presents the distribution of intrinsic absolute UV magnitudes as a function of redshift for the full photometrically selected galaxy sample spanning $z=6$ to $z=16$. Thanks to gravitational lensing, GLIMPSE observations reach nearly three magnitudes deeper than blank-field deep \jwst\ surveys, as indicated by the dashed line.

In its first years of operation, \jwst\ has uncovered a surprisingly high abundance of UV-luminous galaxies at $z > 10$, challenging most theoretical predictions of early galaxy formation \citep{donnan24,bouwens23,mcleod24,austin23,chemerynska24a,adams24}. The two $z\sim16$ candidates found in GLIMPSE \citep{kokorev25} are also consistent with being progenitors of the UV-bright galaxies found at $z\sim12$ when assuming a constant star formation history. This unexpected result has prompted the development of various physical scenarios to explain the apparently elevated star formation efficiency and its slower-than-expected evolution with redshift. Proposed models include stochastic star formation histories \citep[e.g.,][]{pallottini23,sun23b,shen23,munoz23,kravtsov24}, feedback-free starbursts \citep{li23,dekel23}, and dust-free star-forming regions \citep{Ferrara24}. While these models have primarily been designed to reproduce the bright end of the UV luminosity function at $z > 10$, they must also be consistent with observations at fainter magnitudes. In other words, the underlying physical mechanisms must account for the observed properties of both bright and faint galaxies. GLIMPSE observations will provide a powerful complementary test of these competing scenarios.

\subsection{Cosmic Reionization}

The contribution of galaxies to cosmic reionization critically depends on the shape of the UV luminosity function, particularly the number density of the faintest galaxies. Small variations in the faint-end slope of the UV LF can lead to substantial changes in the relative contributions of bright versus faint galaxies. Figure~\ref{fig:alpha} illustrates how the fractional contribution of galaxies in different luminosity bins varies with the faint-end slope $\alpha$. For a slope of $\alpha = -1.8$, galaxies fainter than $M_{\mathrm{UV}} = -16$ contribute approximately 40\% of the total UV luminosity density, comparable to the contribution from galaxies brighter than $M_{\mathrm{UV}} = -18$. In contrast, a steeper slope of $\alpha = -2.1$ results in galaxies fainter than $M_{\mathrm{UV}} = -16$ contributing up to 60\% of the total UV luminosity budget. Using the flux excess in the medium bands, GLIMPSE also enables us to constrain the [O {\sc iii}] + H$\beta$ luminosity function down to $10^{39}$erg s$^{-1}$ at $z=7-9$ \citep{korber25}, accounting for the bulk of star formation at $z=7-9$. The resulting LF has a faint-end slope of $\alpha\approx -1.55\text{ to }-1.78$ at $z=7-9$, which is significantly flatter than the typical UV LF slope derived with the \glimpse\ data \citep{atek26} and in the literature \citep[$\alpha\leq-2$,][]{atek18,bouwens22b}. This implies a reduced contribution of the faintest galaxies to the cosmic reionization \citep{korber25}, lowering the enhanced reionization photon budget implied by earlier JWST observations \citep{munoz24, Atek24a}.

Beyond increasing the census of faint galaxies at $z > 6$, it is essential to estimate their ionizing photon output. The inclusion of medium-band filters enables the isolation of the \ha\ emission line, allowing us to infer the \ha\ flux from the photometric excess in the F480M filter. When combined with measurements of the rest-frame UV emission, this approach facilitates the derivation of the ionizing photon production efficiency, \xiion, for faint galaxies in the redshift range $z = 6.0-6.7$ (Chisholm et al., in prep). \xiion\ is a measure of the ionizing photon production $\dot{N}_{\rm ion}$, which is derived from a hydrogen recombination emission line such as \ha\, and divided by the UV continuum luminosity density at 1500\,\AA.  

Ultimately, only a fraction of these ionizing photons, characterized by the Lyman continuum escape fraction \fesc, can escape the interstellar medium of galaxies and contribute to the ionization of the intergalactic medium. Direct measurements of \fesc\ are not possible during the epoch of reionization due to the high opacity of the IGM. However, indirect indicators calibrated at lower redshifts can be employed \citep[e.g.,][]{flury22b}. For instance, our deep NIRCam imaging allows us to measure the UV continuum slope $\beta$, which has been shown to correlate with \fesc\ \citep[e.g.,][]{chisholm22}. These observations will provide indirect estimates of the \fesc\ in ultra-faint galaxies in the early universe (Jecmen et al., in prep). 

\begin{figure*}
  \centering
    \includegraphics[width=0.8\linewidth]{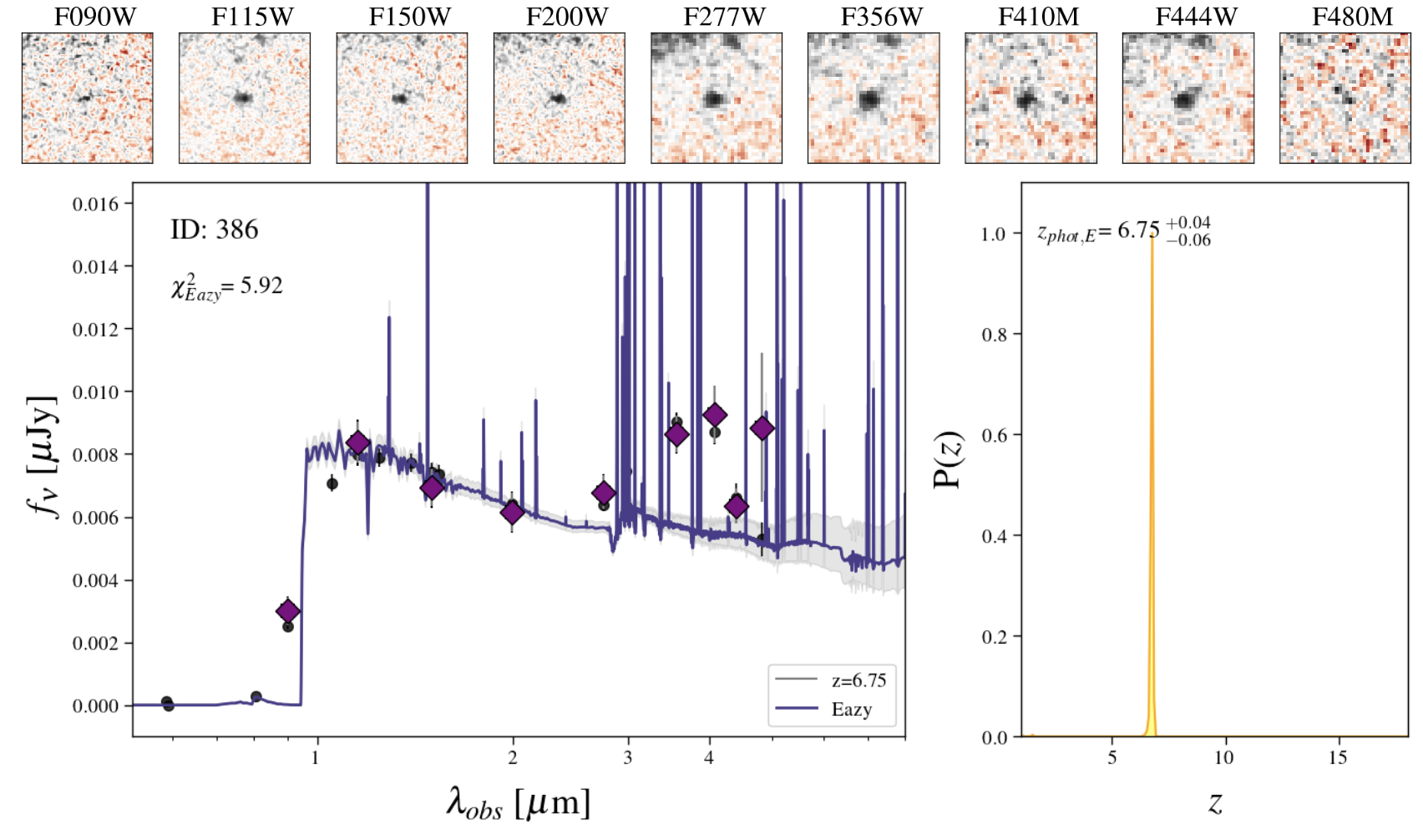}\\
    \vspace{5mm}
        \includegraphics[width=0.8\linewidth]{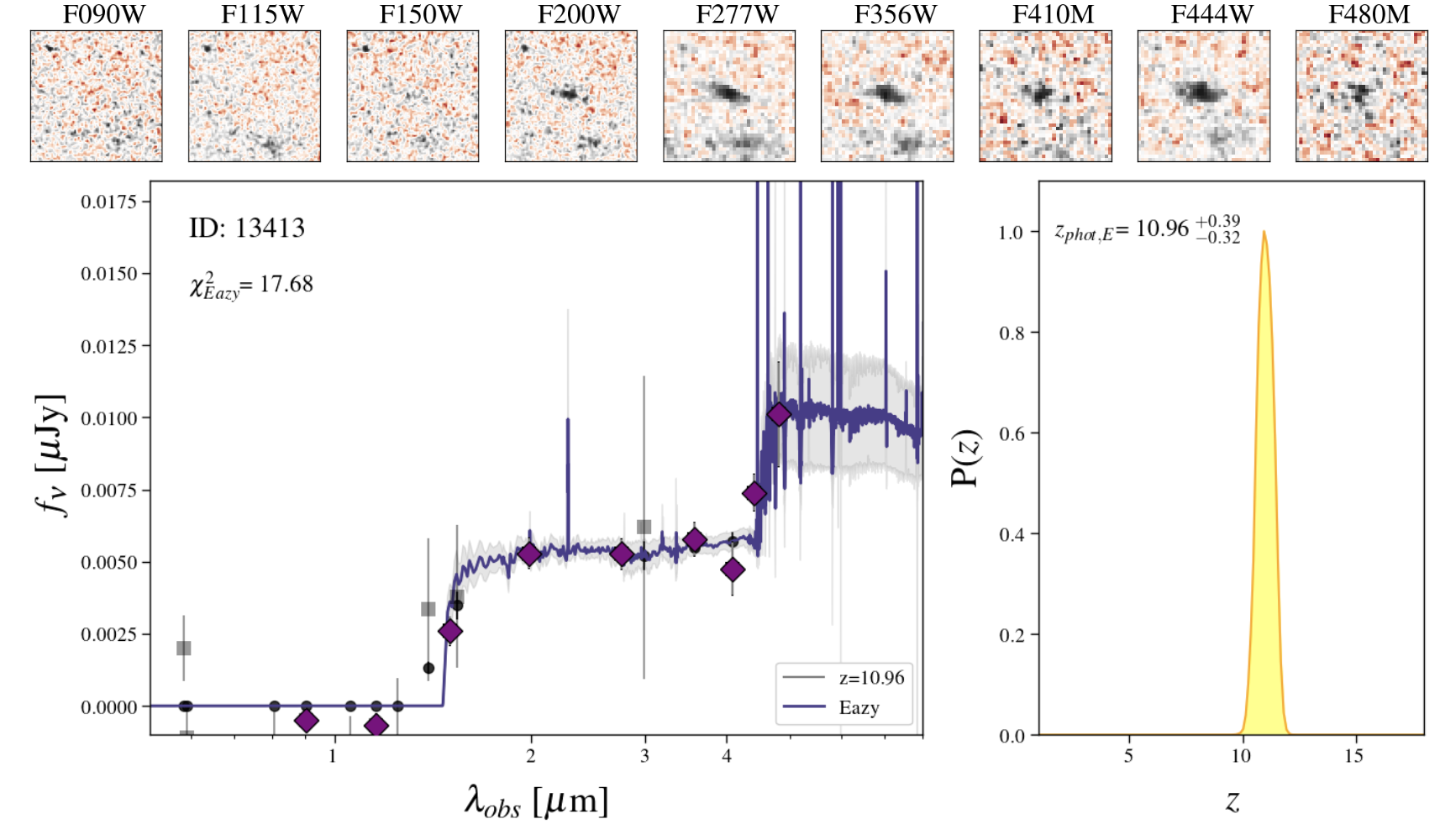}\\
        \caption{Examples of galaxies selected at redshift $z > 6$ using GLIMPSE photometric data. The top and bottom panels show galaxies with best-fit photometric redshifts of $z = 6.75$ and $z = 10.96$, respectively. For each galaxy, the top row presents imaging in the 9 NIRCam filters. The bottom-left panel displays the observed photometry (purple diamonds for \glimpse\ data and gray squares for ancillary data) and the corresponding best-fit spectral energy distribution (blue curve) with associated uncertainties (grey-shaded region with associated best-fit redshift in the legend). The bottom-right panel shows the posterior probability distribution of the photometric redshift.}

    \label{fig:cutouts1}
\end{figure*}

\begin{figure*}
  \centering
        \includegraphics[width=0.8\linewidth]{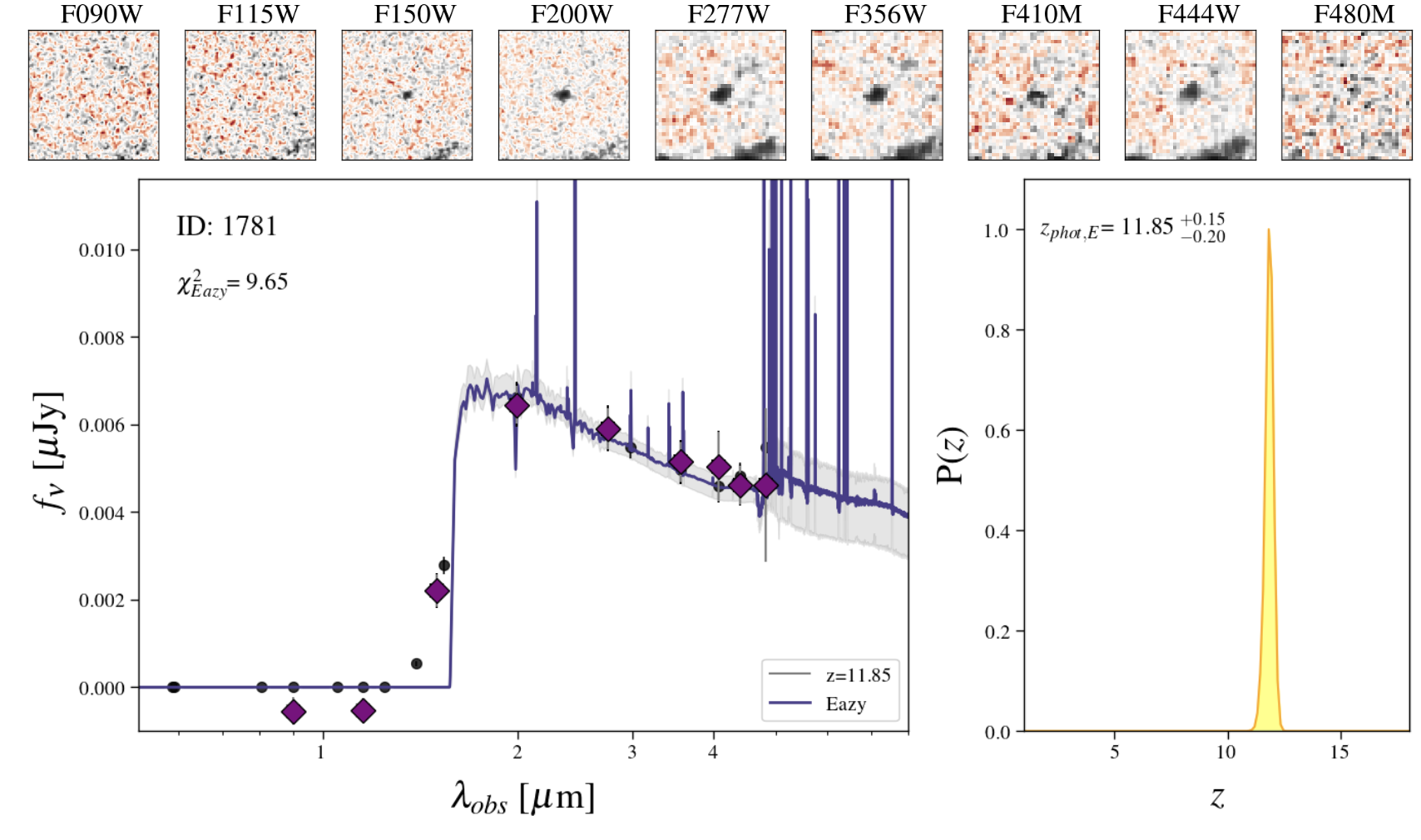}
        \caption{Same as Figure \ref{fig:cutouts1} for a galaxy at $z \sim 12$. }
    \label{fig:cutouts2}
\end{figure*}

\begin{figure}
    \centering
    \includegraphics[width=\linewidth]{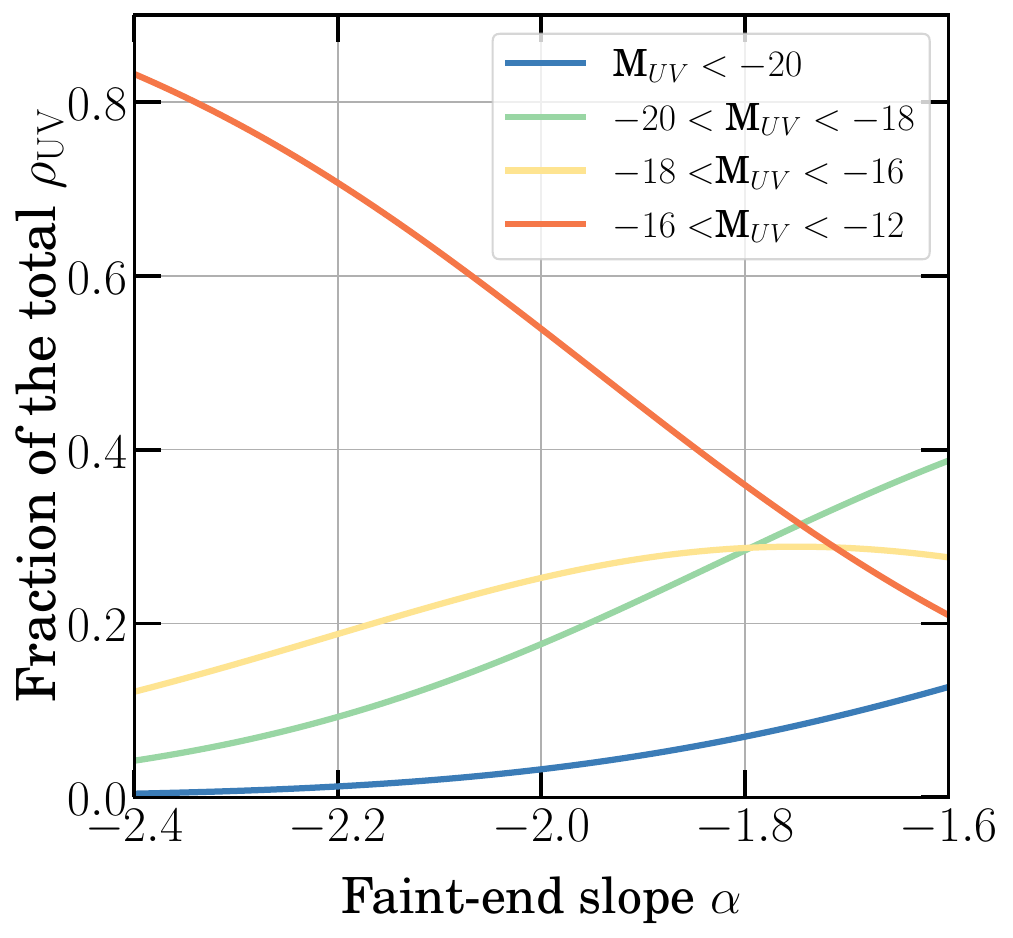}
    \caption{The contribution of galaxies from different ranges of luminosities to the total UV luminosity density (integrated down to \muv $=-12$) as a function of the faint-end slope $\alpha$ of the UV luminosity function.}
    \label{fig:alpha}
\end{figure}

\subsection{Resolved star-formation processes in high$-z$ galaxies}

Strong gravitational lensing enables spatially resolved studies of distant galaxies at unprecedented detail. \hst\ observations of the AS1063 lensing cluster revealed some of the faintest and most compact sources at high-redshift, possibly progenitors of globular clusters \citep{vanzella17}. Beyond resolving global galaxy structures and star-forming regions, the GLIMPSE observations allow us to identify and characterize individual star-forming clumps and star clusters within high-redshift galaxies \citep[e.g.,][]{claeyssens23,claeyssens25}. Investigating the properties of these clumps and their connection to galaxy mass assembly and evolution offers critical insight into the processes governing star formation on small physical scales, down to $\sim$100 pc or below. Recent \jwst\ observations of lensing clusters have demonstrated the power of gravitational lensing, at very high magnifications, in probing stellar substructures with typical sizes of individual stellar clusters, some of which exhibit properties consistent with globular clusters \citep[e.g.,][]{mowla22, claeyssens23,vanzella23,adamo24}.
GLIMPSE extends these capabilities by enabling the study of faint stellar clumps, and especially star clusters, with masses below 10$^5$~\msol\ and a spatial resolution approaching $\sim$10 pc. In our analysis, we identified 451 individual stellar clumps, from 61 highly magnified galaxies (with gravitational lensing magnification factors spanning $\mu = 5$ to $\mu = 140$) across the redshift range $z = 0.5$-$8.2$ (Claeyssens et al., in prep). The depth of the GLIMPSE observations allows us to detect extremely faint systems, reaching absolute UV magnitudes as low as \muv\ $\approx -8.5$. Among these, 145 clumps exhibit compact morphologies with typical sizes smaller than 20 pc, thus being consistent with individual star clusters which represent the largest collection of such systems to date.

\begin{figure*}
    \includegraphics[width=0.34\textwidth]{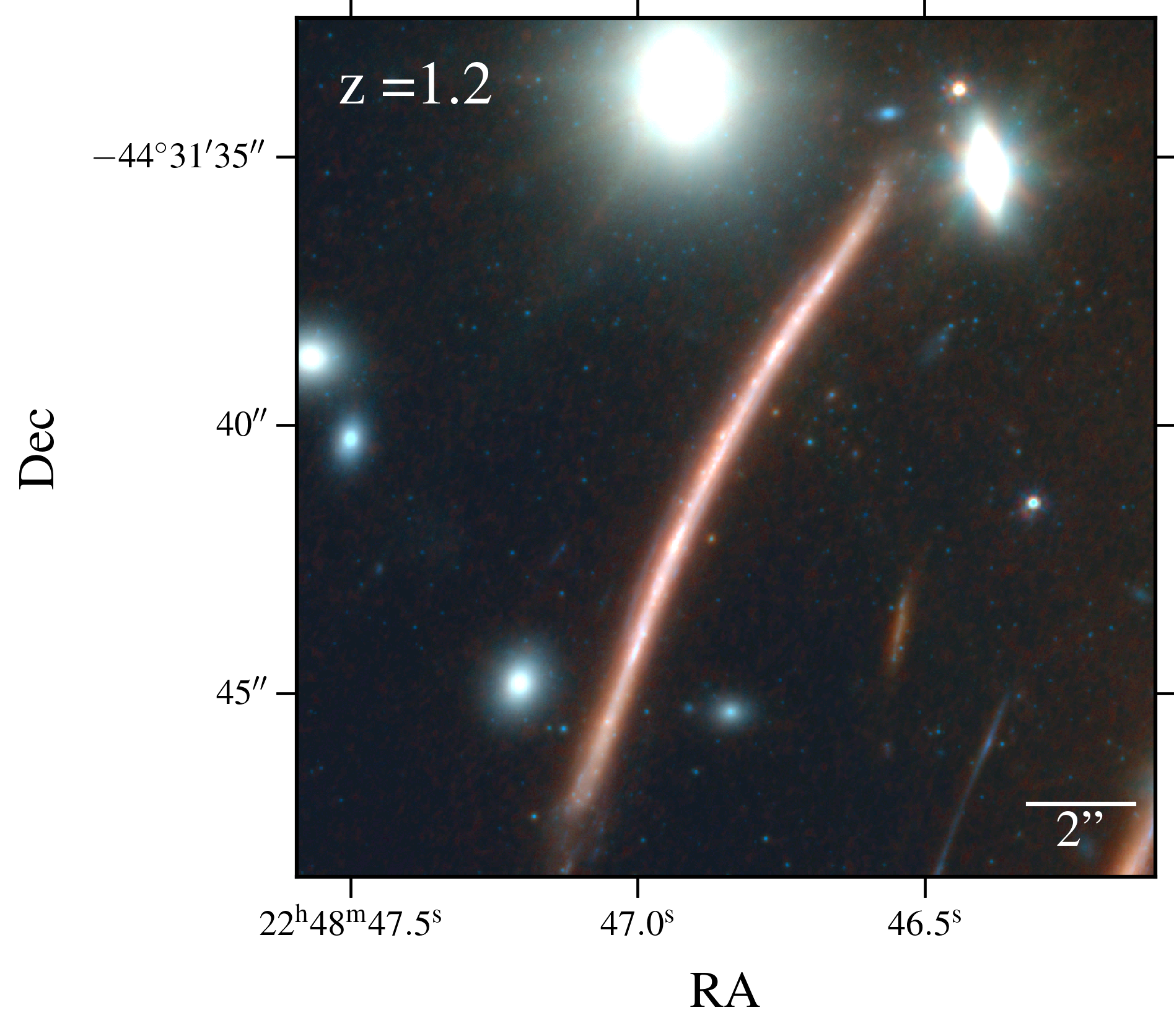}
    \hspace{-0.3cm}
    \includegraphics[width=0.34\textwidth]{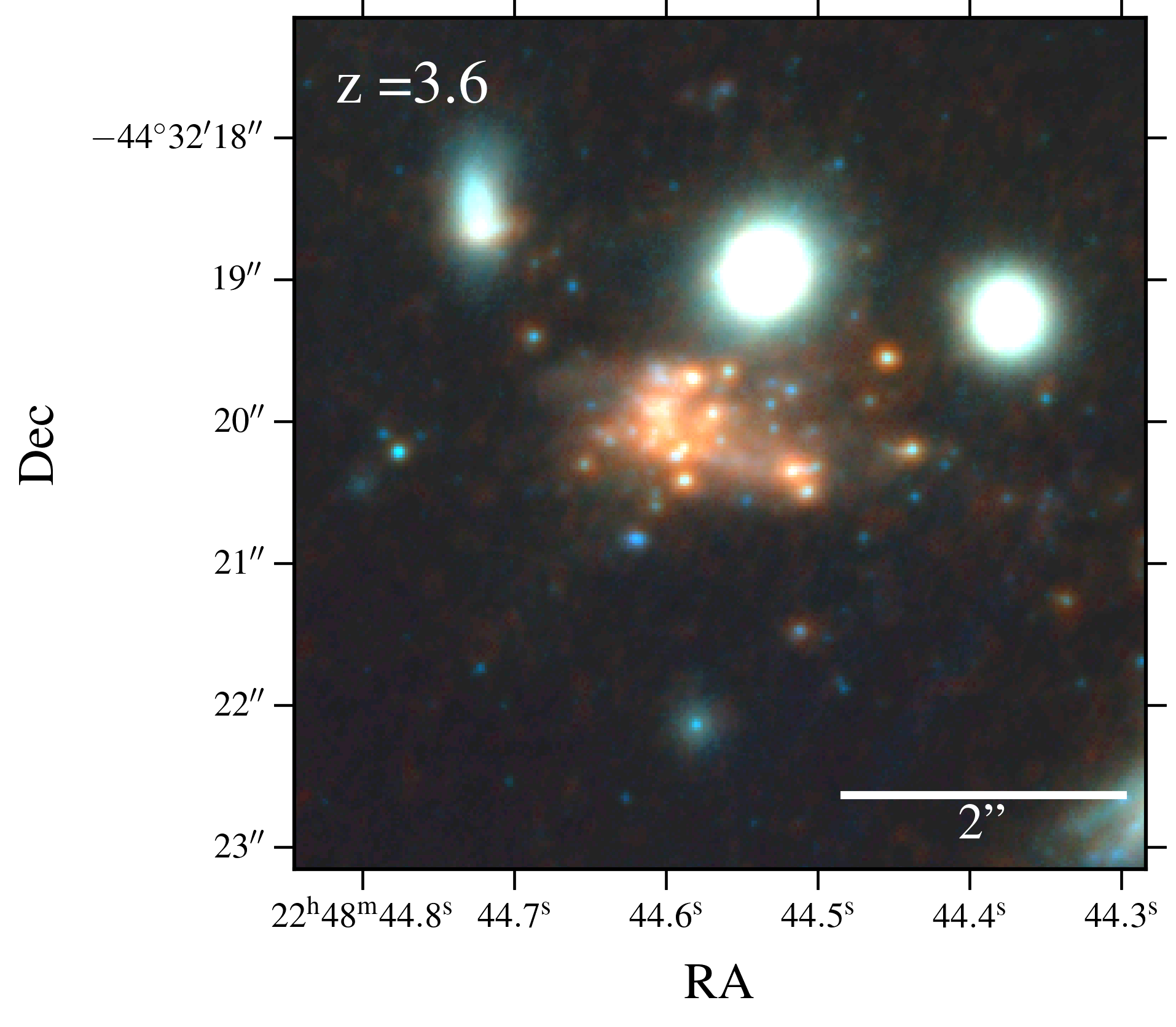}
    \hspace{-0.3cm}
    \includegraphics[width=0.34\textwidth]{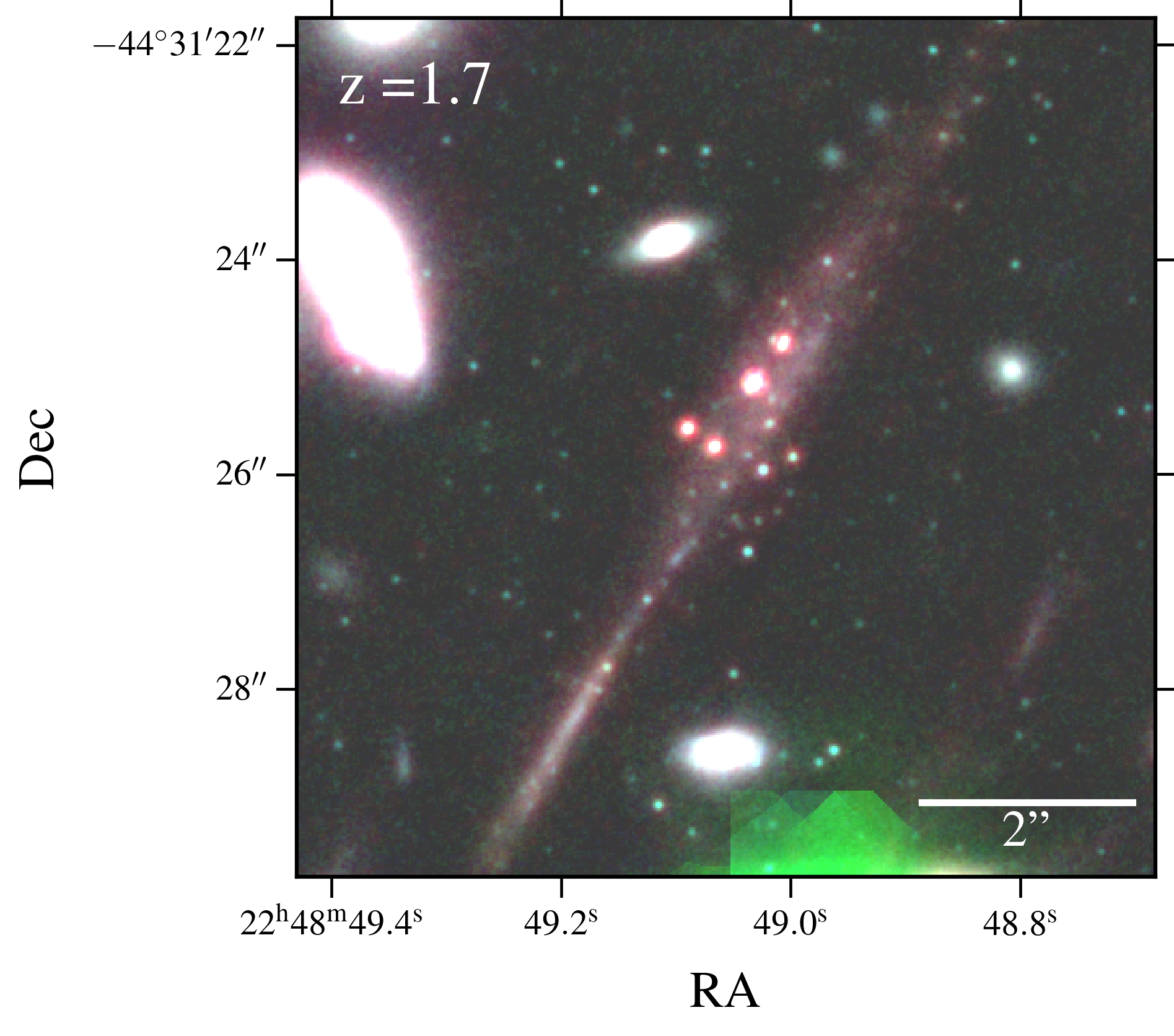}\\
   \includegraphics[width=0.34\textwidth]{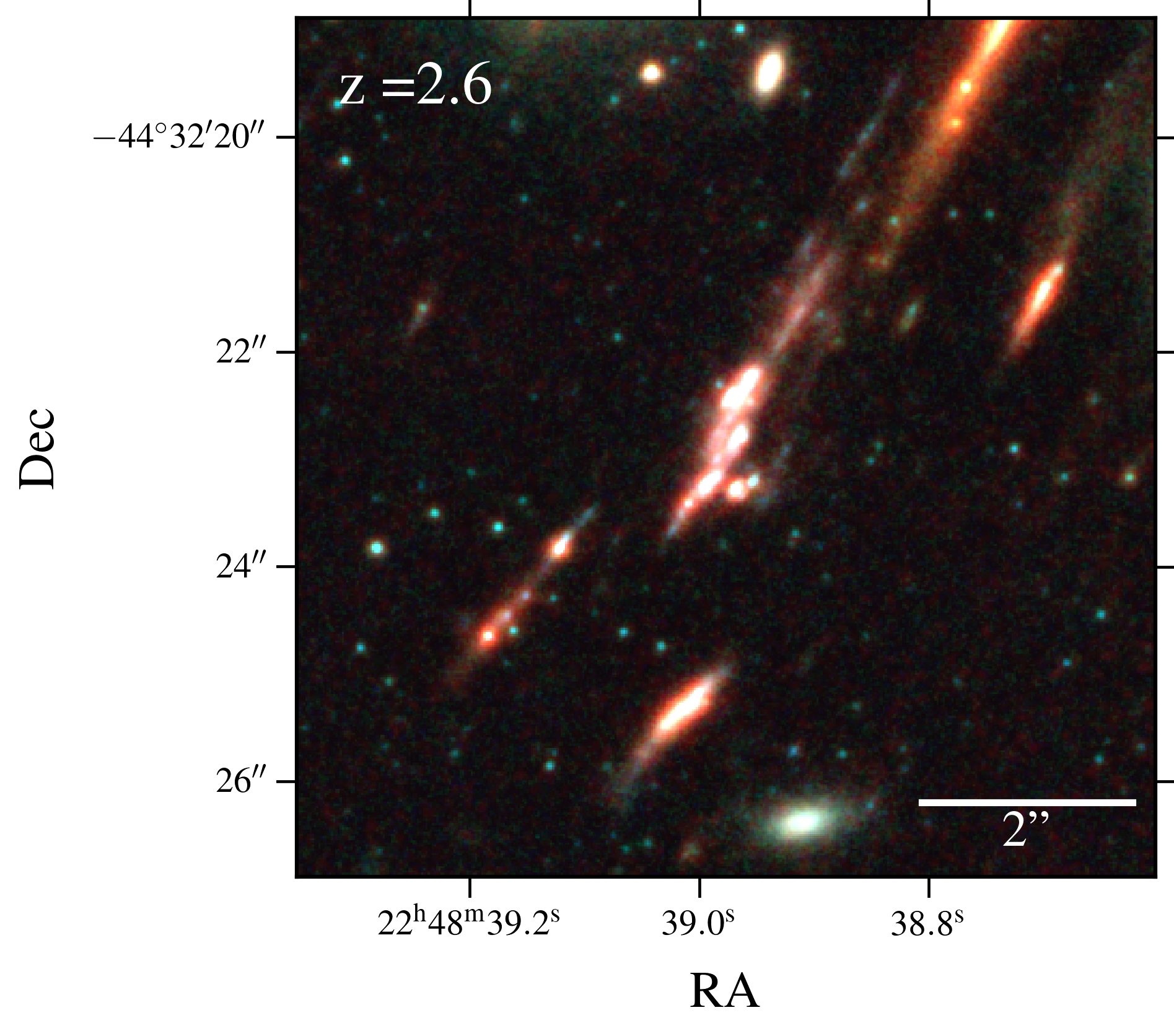}
    \hspace{-0.3cm}
    \includegraphics[width=0.34\textwidth]{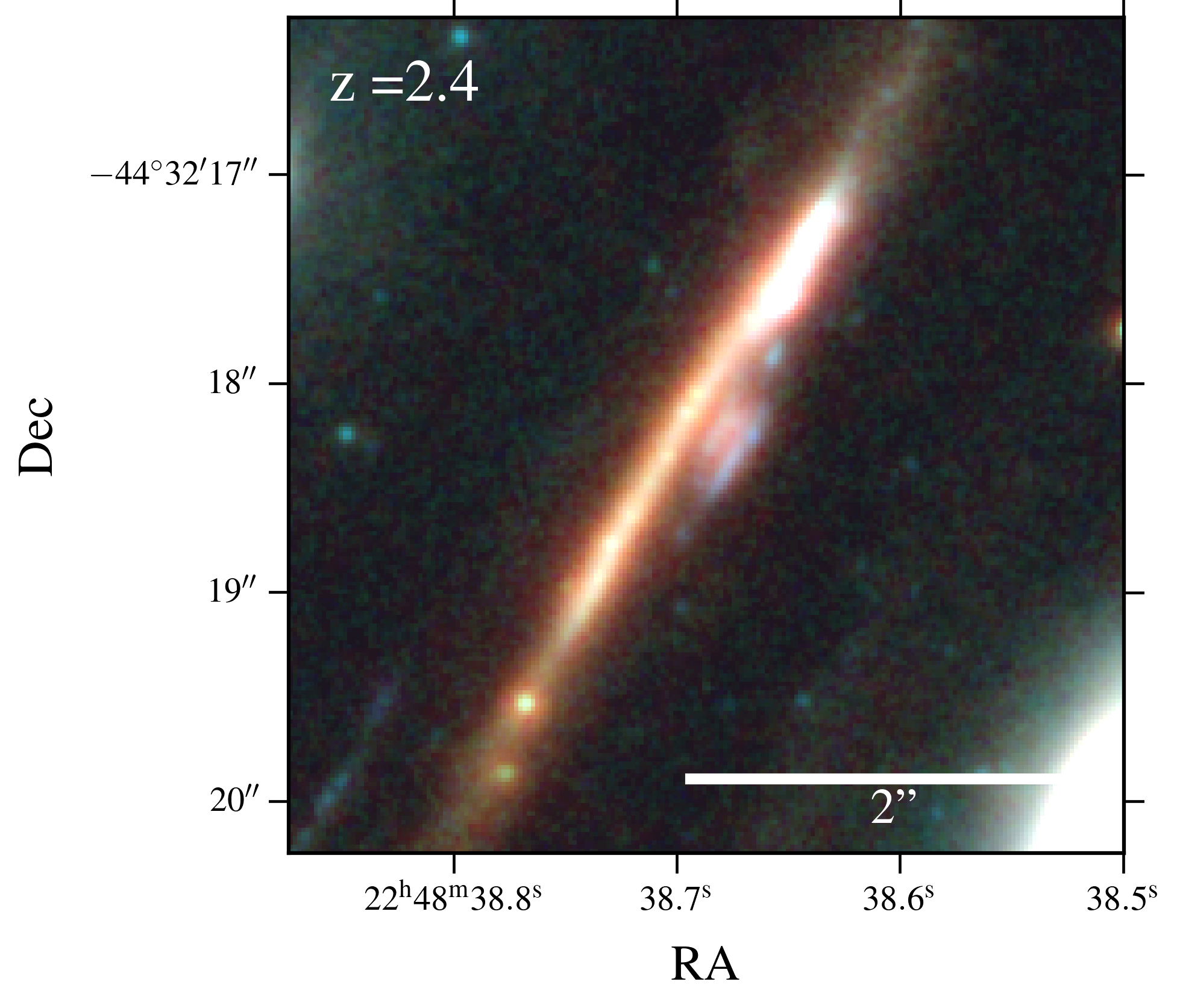}
    \hspace{-0.3cm}
    \includegraphics[width=0.34\textwidth]{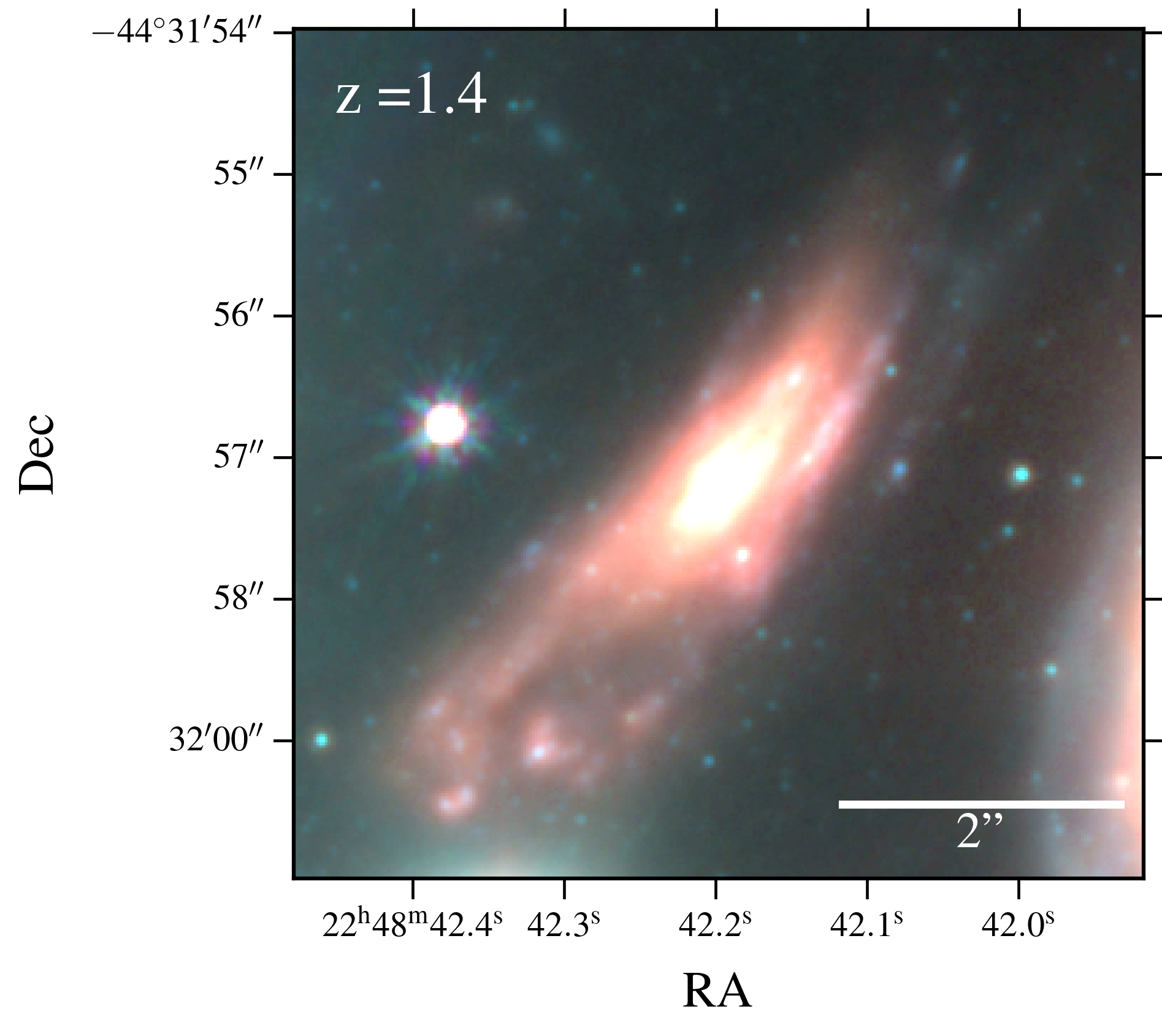}
     \caption{Examples of strongly lensed galaxies at various redshifts. The color image stamps showcase a wide variety of morphologies among spatially-resolved systems, including bright star-forming clumps and compact star clusters. The depth and lensing magnification enables the identification ultra-faint stellar clumps down to \muv=$-8.5$ mag, stellar masses below $10^{5}$\msol, and a spatial resolution down to $\sim10$ pc. Each panel is labeled with the galaxy redshift and a physical scale. The green feature visible in the top right panel is due to bCG residuals and the scaling that has been chosen to better display the clumpy structures}
    \label{fig:clumps}
\end{figure*}

\subsection{Other lensing-enabled science}

Another important research avenue enabled by strong gravitational lensing is the discovery of individual stars in distant galaxies. When stars lie near or cross the caustic lines of a lensing cluster, they can be magnified by factors of several thousands, becoming sufficiently bright to be detected by current instrumentation \citep{miralda91}. In recent years, several such highly magnified stars have been identified at record-breaking redshifts, from $z \sim 1$ to $z \sim 6$, through deep \hst\ and \jwst\ imaging of lensing clusters \citep{kelly18,welch22,furtak24,fudamoto25}. Follow-up \jwst\ spectroscopy has provided unprecedented constraints on their physical properties, including mass, temperature, and composition \citep[e.g.,][]{meena23}. These observations offer unique insights into the IMF, stellar evolution, and chemical enrichment in the early Universe, providing direct constraints on the stellar populations of young galaxies.

Thanks to its exceptional depth, GLIMPSE enables, for the first time, the detection of lower-mass individual stars at high redshift. We have identified two caustic-crossing events at redshifts $z \sim 1.8$ and $z \sim 4.3$. The photometric properties of these point sources are best matched by asymptotic giant branch (AGB) star templates (Furtak et al., in prep), suggesting the feasibility of studying evolved stellar populations at cosmological distances.

Similarly, strong gravitational lensing combined with ultra-deep \jwst\ imaging provides a powerful avenue to pinpoint the first generation of stars, the so-called Population~III (Pop~III) stars. These primordial systems can be isolated by identifying key spectral signatures in photometric observations, including a pronounced Balmer break, strong hydrogen recombination emission, and a lack of oxygen features \citep{fujimoto25}. One of the most promising candidates has been identified at $z\sim6$ behind the lensing cluster A2744, with an amplification factor around $\mu\sim80$ \citep{morishita25}.

\section{Summary}

In this paper, we present the basic properties and scientific goals of the \jwst\ GLIMPSE survey, which constitutes the deepest exploration of the distant Universe to date. By combining ultra-deep NIRCam imaging with the strong gravitational lensing of the Abell S1063 cluster, GLIMPSE achieves a 5$\sigma$ depth of AB $\sim$ 31. Despite its limited survey area, the intrinsic (delensed) depth reaches beyond AB = 33 mag. The imaging data consists of seven broadband filters from F090W to F444W, and two medium-band filters, F410M and F480M. This provides a unique opportunity to detect the faintest sources from Cosmic Dawn to Cosmic Reionization, enable spatially resolved studies of high-redshift star-forming galaxies, and uncover unique lensed objects such as individual stars. 

The GLIMPSE survey enables a broad range of scientific investigations, which we summarize as follows:

\begin{itemize}
    \item Probing the extreme faint end of the UV luminosity function from $z \sim 6$ to $z \sim 16$. While the first years of \jwst\ observations have constrained the bright end of the UV LF, revealing a surprisingly high number of UV-bright galaxies, the faint population has largely remained unexplored. The precise shape of the faint end and the abundance of UV-faint galaxies provide stringent tests of galaxy formation models, which have never been probed in this luminosity regime. Furthermore, these measurements offer an additional benchmark for models attempting to explain the overabundance of UV-bright galaxies at $z > 9$. We have identified nearly 540 galaxies at $z > 6$, including 411 sources at $6 < z < 9$ with UV magnitudes spanning \muv$= [-20, -12]$, 125 sources at $9 < z < 15$, and 2 sources at $z \sim 16$. This sample provides, for the first time, a glimpse into the faint galaxy population at these early cosmic epochs.
    \item At the same time, probing the faintest galaxies allows for a more complete census of galaxies during the epoch of reionization. In addition, we can measure the ionizing photon production of faint galaxies in the redshift range $z = 6-6.7$ using the \ha\ flux excess in the F480M medium band.
    \item Strong lensing also provides unique spatial information on distant galaxies, including star clusters with sizes below 20 pc and masses down to $10^{5}$ \msol, offering valuable insight into the mass assembly and evolution of galaxies. Along the critical lines of the lens, objects can be magnified by factors of several thousand. The GLIMPSE data have revealed two candidate low-mass individual stars at $z \sim 1.8$ and $z \sim 4.3$. 
\end{itemize}

This paper is accompanied by the first data release (DR1) of the GLIMPSE survey \footnote{\url{https://jwst-glimpse.github.io}} \footnote{\url{https://zenodo.org/records/22552278}}. The release includes all reduced \jwst\ NIRCam mosaics at 20 mas~pix$^{-1}$ and 40 mas~pix$^{-1}$ resolution for the SW and LW filters, respectively. We also provide reduced and astrometrically aligned \hst\ mosaics combining data from multiple programs. In addition, we publish the best strong-lensing model, based on constraints from both the HFF and GLIMPSE datasets. Finally, the release contains photometric catalogs with flux measurements in circular apertures of varying diameters, ranging from $D = 0\farcs{1}$ to $1\farcs{2}$.

\section*{Acknowledgments}

HA and IC acknowledge support from CNES, focused on the JWST mission, and the Programme National Cosmology and Galaxies (PNCG) of CNRS/INSU with INP and IN2P3, co-funded by CEA and CNES. IC acknowledges funding support from the Initiative Physique des Infinis (IPI), a research training program of the Idex SUPER at Sorbonne Universit\'e. HA acknowledges support by the French National Research Agency (ANR) under grant ANR-21-CE31-0838. This work has made use of the \texttt{CANDIDE} Cluster at the \textit{Institut d'Astrophysique de Paris} (IAP), made possible by grants from the PNCG and the region of Île de France through the program DIM-ACAV+, and the Cosmic Dawn Center and maintained by S. Rouberol. PO has received funding from the Swiss State Secretariat for Education, Research and Innovation (SERI) under contract number MB22.00072, as well as from the Swiss National Science Foundation (SNSF) through project grant 200020\_207349. The Cosmic Dawn Center (DAWN) is funded by the Danish National Research Foundation under grant DNRF140. PN gratefully acknowledges funding from the Department of Energy grant DE-SC0017660 and grant 63406 from the John Templeton Foundation. AZ acknowledges support by the Israel Science Foundation Grant No. 864/23. AA acknowledges support by the Swedish research council Vetenskapsr{\aa}det (VR 2021-05559, and VR consolidator grant 2024-02061).

This work is based on observations obtained with the NASA/ESA/CSA \textit{JWST} and the NASA/ESA \textit{Hubble Space Telescope} (HST), retrieved from the \texttt{Mikulski Archive for Space Telescopes} (\texttt{MAST}) at the \textit{Space Telescope Science Institute} (STScI). STScI is operated by the Association of Universities for Research in Astronomy, Inc. under NASA contract NAS 5-26555. The data described here may be obtained from
\url{https://doi.org/10.17909/4byn-fe55}.
Support for US investigators in program \#3293 was provided by NASA through a grant from the Space Telescope Science Institute, which is operated by the Association of Universities for Research in Astronomy, Inc., under NASA contract NAS 5-03127.
The Dunlap Institute is funded through an endowment established by the David Dunlap family and the University of Toronto.
We acknowledge the support of the Canadian Space Agency (CSA) [25JWGO4A06].

\appendix

\begin{figure*}
    \centering
    \includegraphics[width=0.32\linewidth]{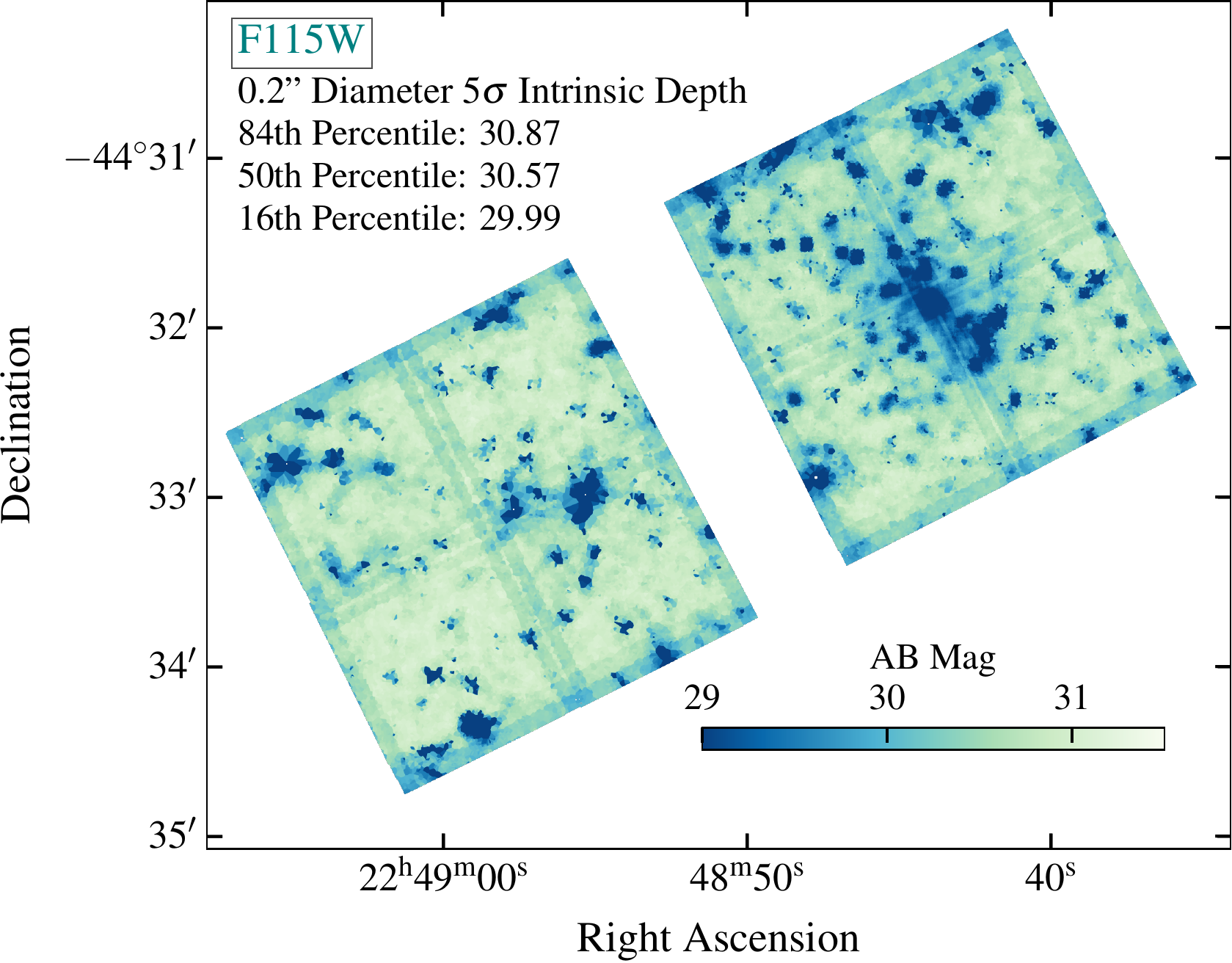}
    \includegraphics[width=0.32\linewidth]{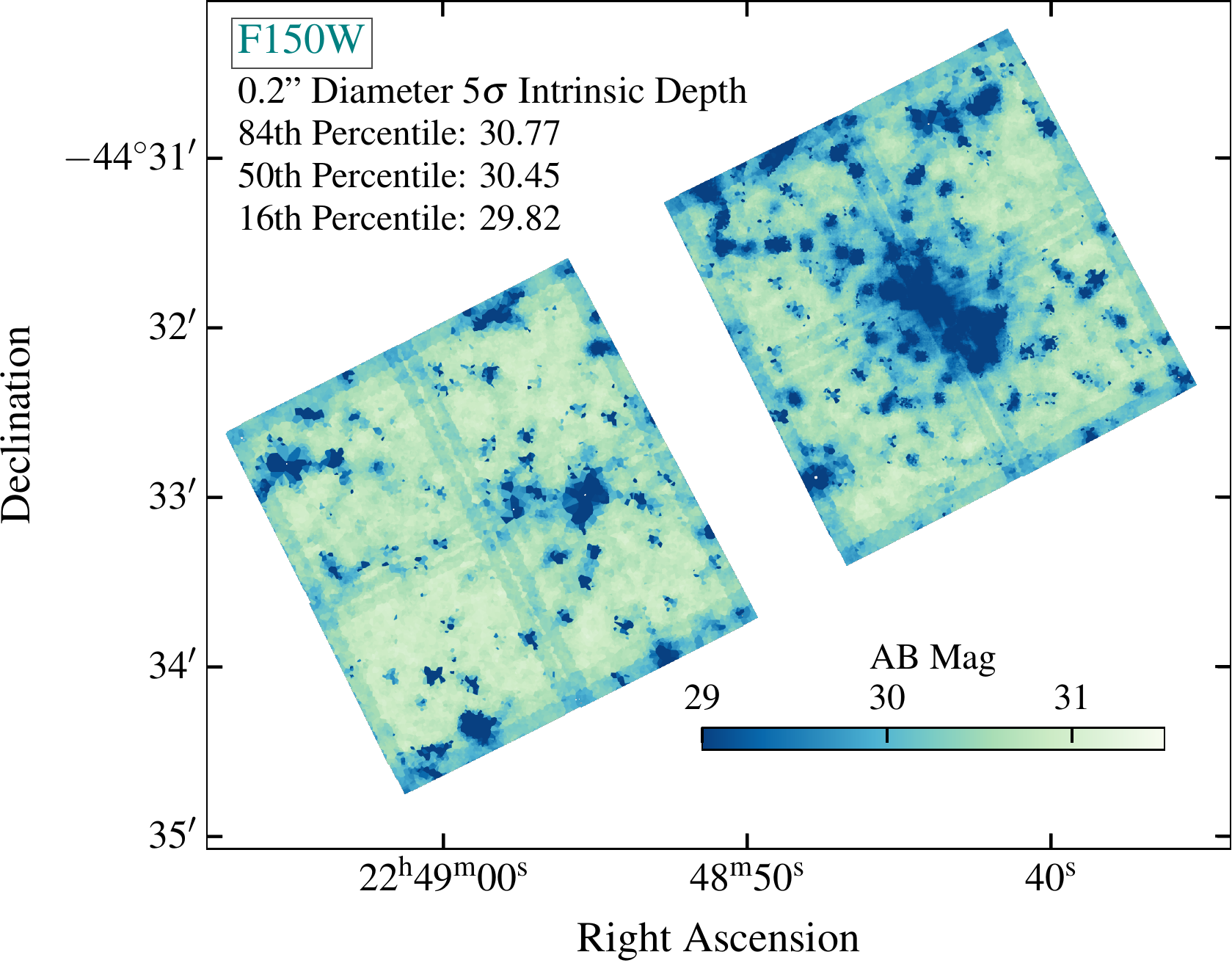}
    \includegraphics[width=0.32\linewidth]{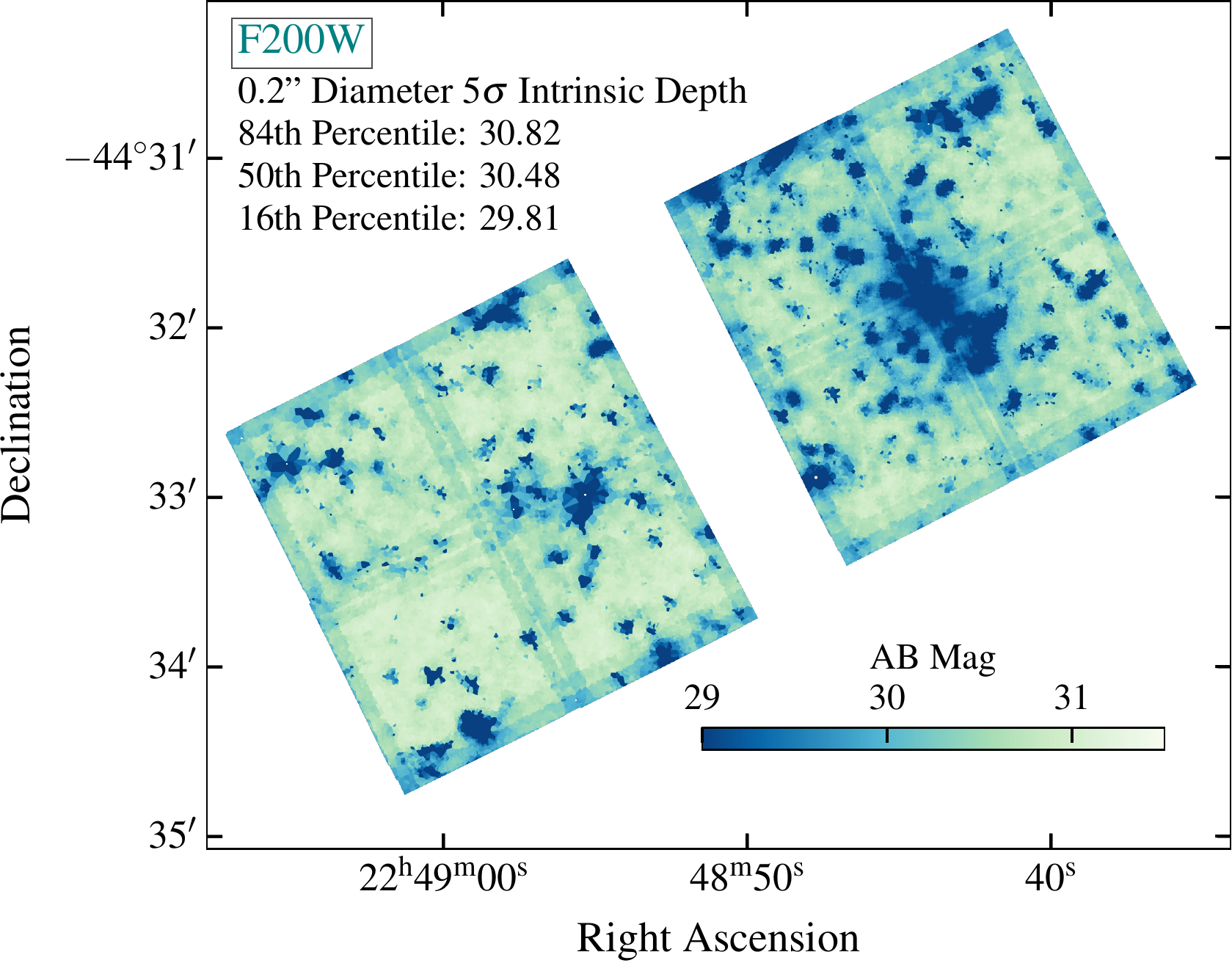}\\
    \vspace{0.4cm}
 \includegraphics[width=0.32\linewidth]{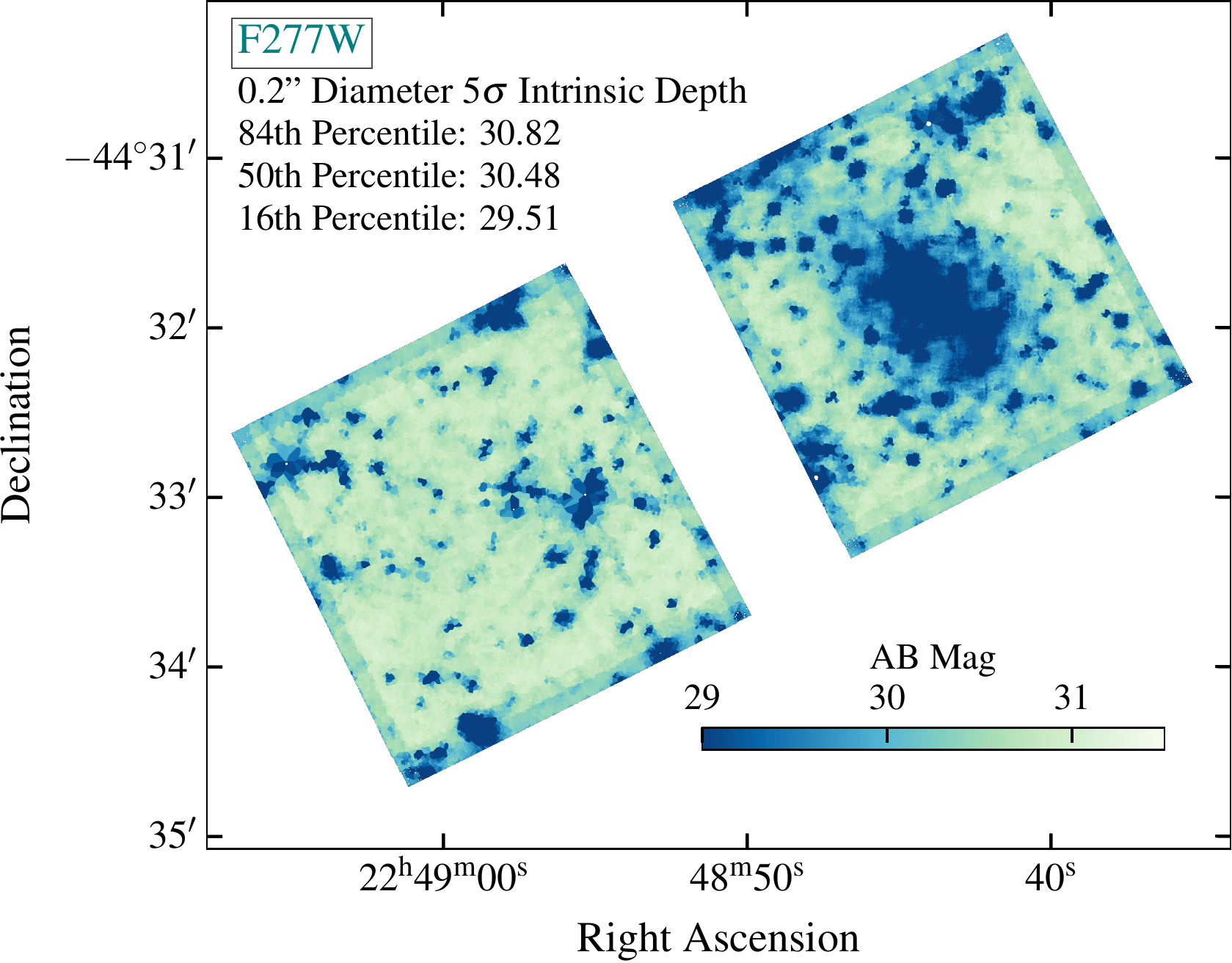}
 \includegraphics[width=0.32\linewidth]{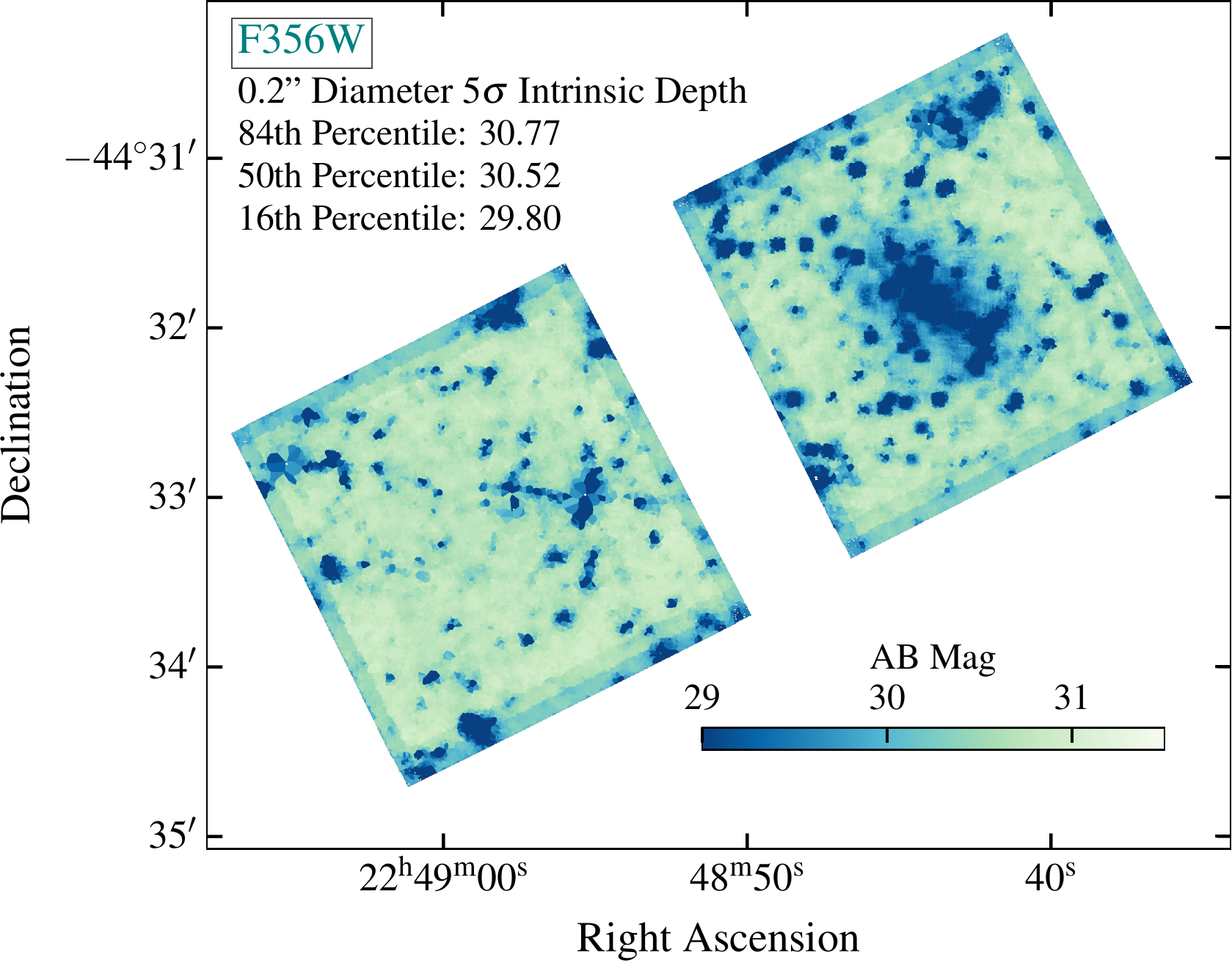}
 \includegraphics[width=0.32\linewidth]{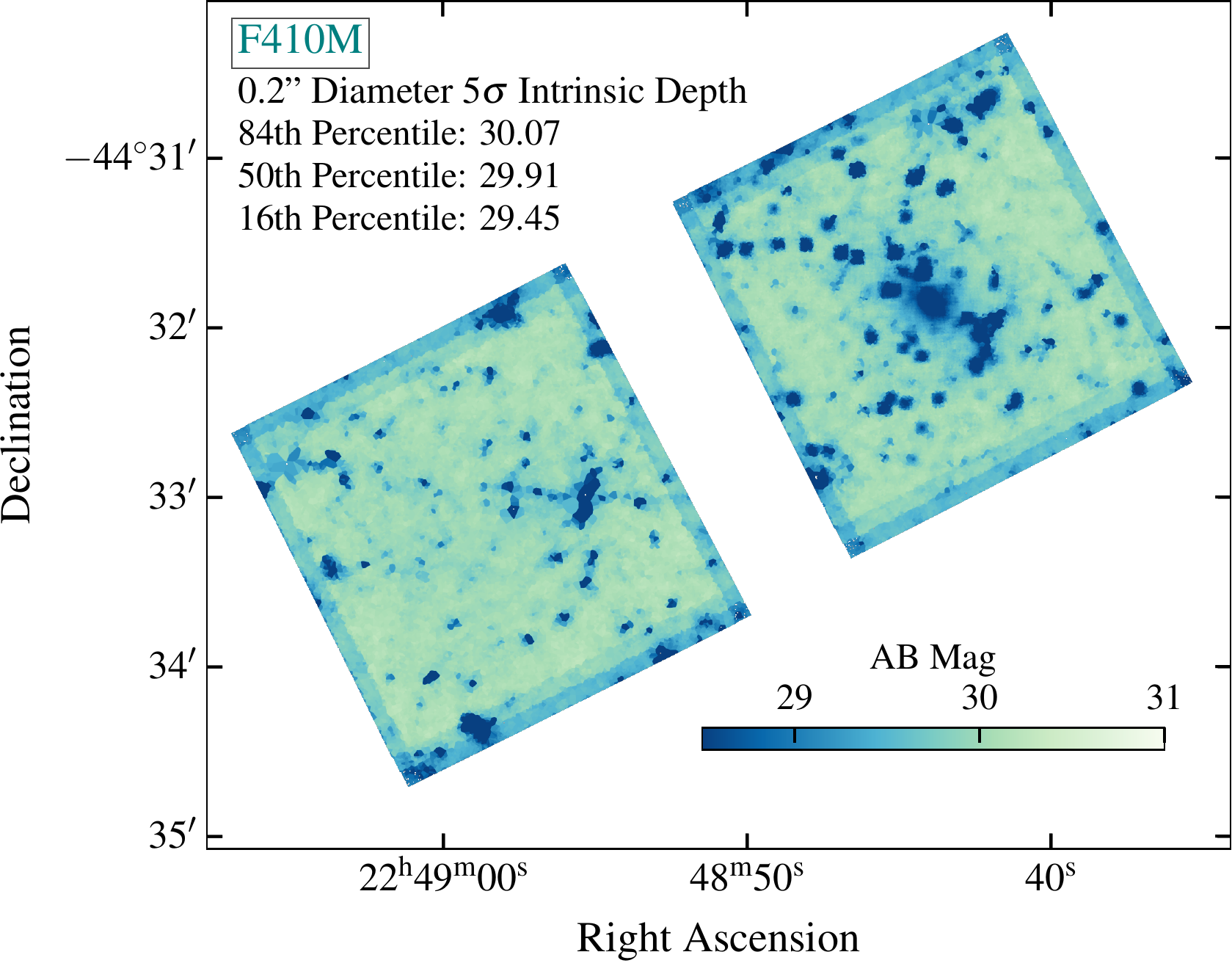}\\
 \vspace{0.4cm}
 \includegraphics[width=0.32\linewidth]{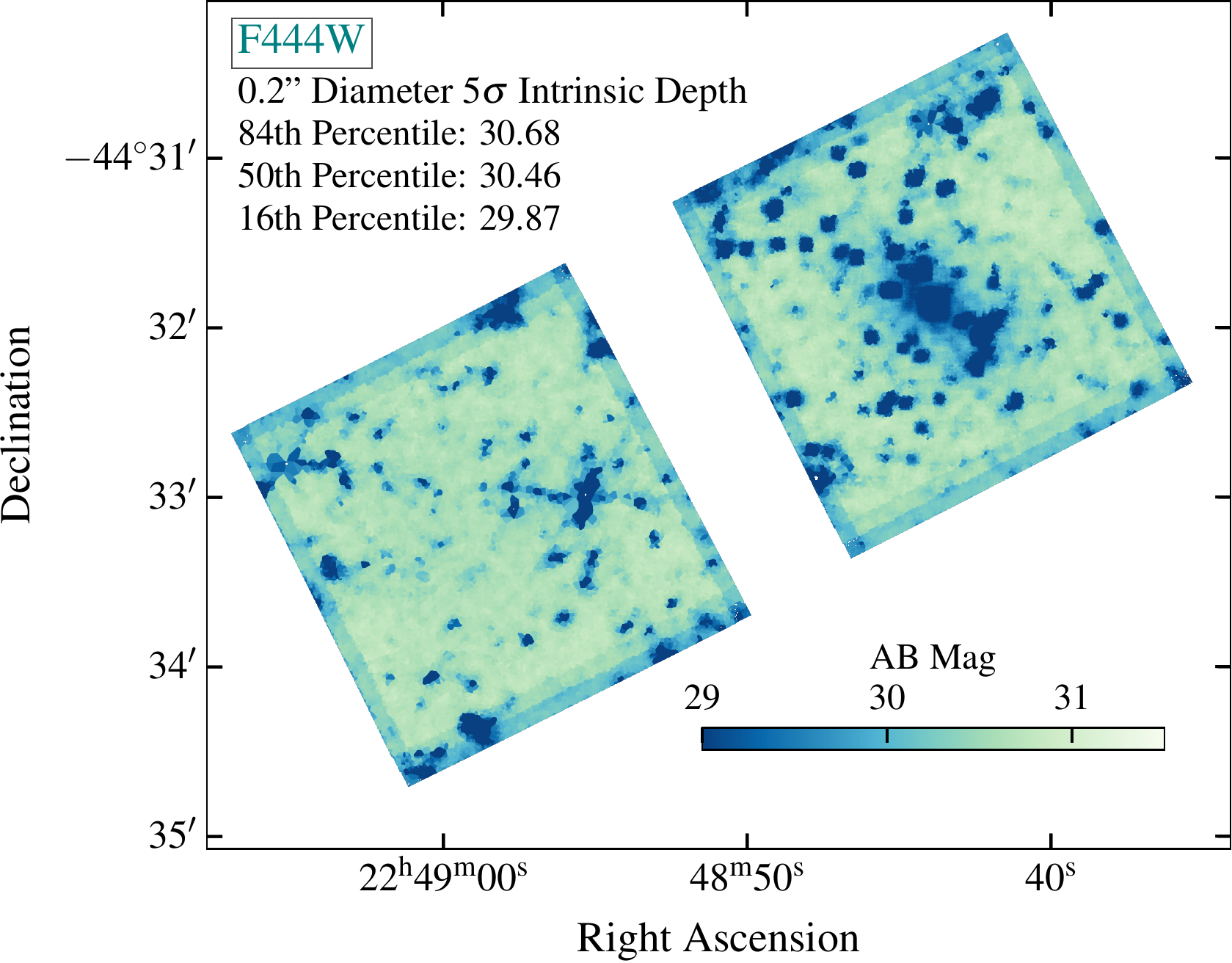}
 \includegraphics[width=0.32\linewidth]{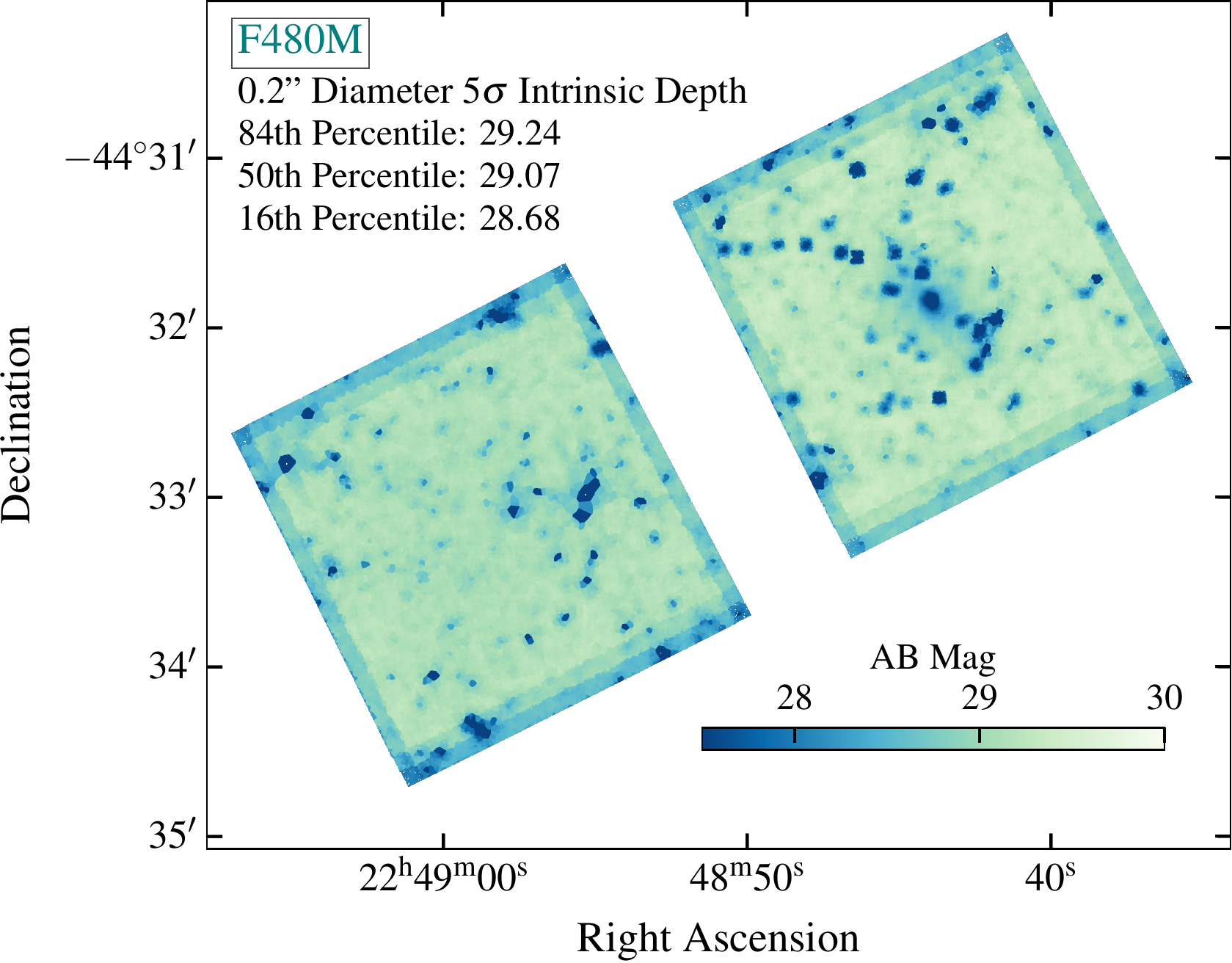}      
    \caption{Depth maps for the rest of the GLIMPSE filters based on the same procedure used for Figure \ref{fig:depth}.}
    \label{fig:placeholder}
\end{figure*}

\bibliography{references}
\bibliographystyle{mnras}

\end{document}